\documentstyle[sprocl,epsfig]{article}

\def\be{\begin{equation}}
\def\bea{\begin{eqnarray}}
\def\eea{\end{eqnarray}}
\newcommand{\mH}{\mbox{$m_{\mathrm{H}}$}}

\newcommand{\mh}{\mbox{$m_{\mathrm{h}}$}}
\newcommand{\mA}{\mbox{$m_{\mathrm{A}}$}}
\newcommand{\ee}{\mbox{$\mathrm{e}^{+}\mathrm{e}^{-}$}}
\newcommand{\ra}{\mbox{$\rightarrow$}}
\newcommand{\sba}{\mbox{$\sin ^2 (\beta -\alpha)$}}
\newcommand{\tanb}{\mbox{$\tan \beta$}}

\newcommand{\tautau}{\mbox{$\tau^{+}\tau^{-}$}}
\newcommand{\bb}{\mbox{$\mathrm{b} \bar{\mathrm{b}}$}}
\newcommand{\mHpm}{\mbox{$m_{\mathrm{H}^{\pm}}$}}
\newcommand{\Hp}{\mbox{$\mathrm{H}^{+}$}}
\newcommand{\Hm}{\mbox{$\mathrm{H}^{-}$}}
\newcommand{\csbar}{\mbox{$\mathrm{c} \bar{\mathrm{s}}$}}
\newcommand{\cbars}{\mbox{$\bar{\mathrm{c}}\mathrm{s}$}}
\newcommand{\tp}{\mbox{$\tau^+$}}
\newcommand{\tm}{\mbox{$\tau^-$}}
\newcommand{\nubar}{\mbox{$\bar{\nu}$}}

\begin{document}
\begin{titlepage}
\def\thefootnote{\fnsymbol{footnote}}       

\begin{center}
\mbox{ } 

\end{center}
\begin{flushright}
\Large
\mbox{\hspace{10.2cm} hep-ph/0112082} \\
\mbox{\hspace{12.0cm} December 2001}
\end{flushright}
\begin{center}
\vskip 1.0cm
{\Huge\bf
Complete LEP Data: 

Status of Higgs Boson Searches}
\vskip 1cm
{\LARGE\bf Andr\'e Sopczak}\\
\smallskip
\Large Lancaster University

\vskip 2.5cm
\centerline{\Large \bf Abstract}
\end{center}

\vskip 3.5cm
\hspace*{-1cm}
\begin{picture}(0.001,0.001)(0,0)
\put(,0){
\begin{minipage}{16cm}
\Large
\renewcommand{\baselinestretch} {1.2}
The LEP experiments completed data-taking in November 2000.
New preliminary combined results of the four LEP experiments
ALEPH, DELPHI, L3 and OPAL are presented for various
Higgs boson searches.
\renewcommand{\baselinestretch} {1.}

\normalsize
\vspace{10cm}
\begin{center}
{\sl \large
Presented at the Third International Conference on Non-Accelerator
New Physics, NANP--01, Moscow, Russia, June 2001
\vspace{-6cm}
}
\end{center}
\end{minipage}
}
\end{picture}
\vfill

\end{titlepage}


\newpage
\thispagestyle{empty}
\mbox{ }
\newpage
\setcounter{page}{1}

\large
\title{\Large Complete LEP Data: Status of Higgs Boson Searches}

\author{\large Andr\'e Sopczak}

\address{\large Lancaster University \\E-mail: andre.sopczak@cern.ch} 


\maketitle\abstracts{\large
The LEP experiments completed data-taking in November 2000.
New preliminary combined results of the four LEP experiments
ALEPH, DELPHI, L3 and OPAL are presented for various
Higgs boson searches.}

After 11 years of operation the LEP experiments have completed data-taking.
In 2000 the center-of-mass energy was pushed to 209 GeV with
most data taken around 206 GeV.
In the last three years of operation a luminosity of about 687~pb$^{-1}$
was delivered to each experiment, which exceeded expectations.
Despite hints for a Higgs boson discovery around 116~GeV
data-taking was not continued in 2001.

\section{Standard Model Higgs Boson}
\begin{figure}[htb]
\vspace*{0.5cm}
\begin{center}
\epsfig{figure=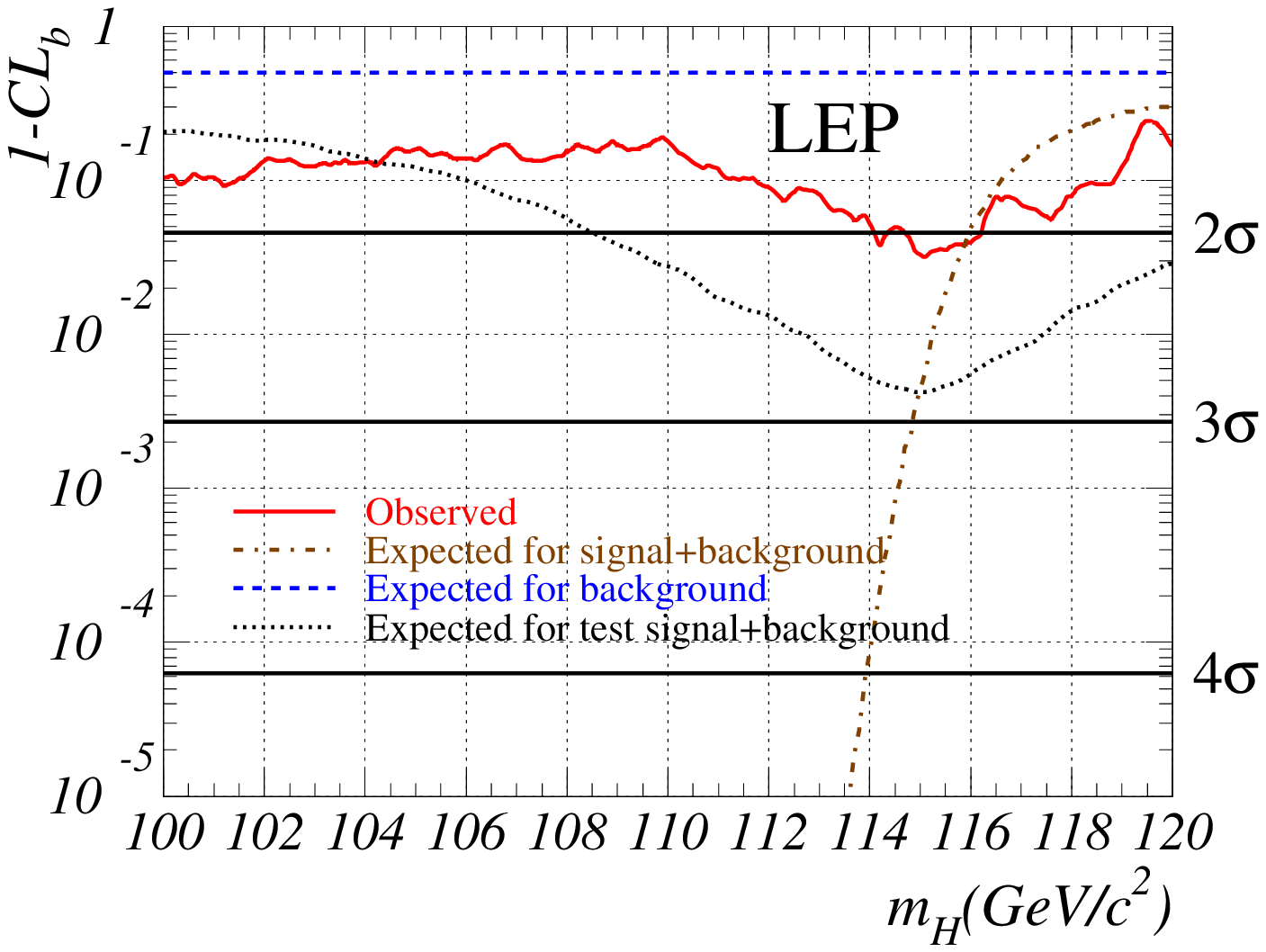,width=0.48\textwidth}
\epsfig{figure=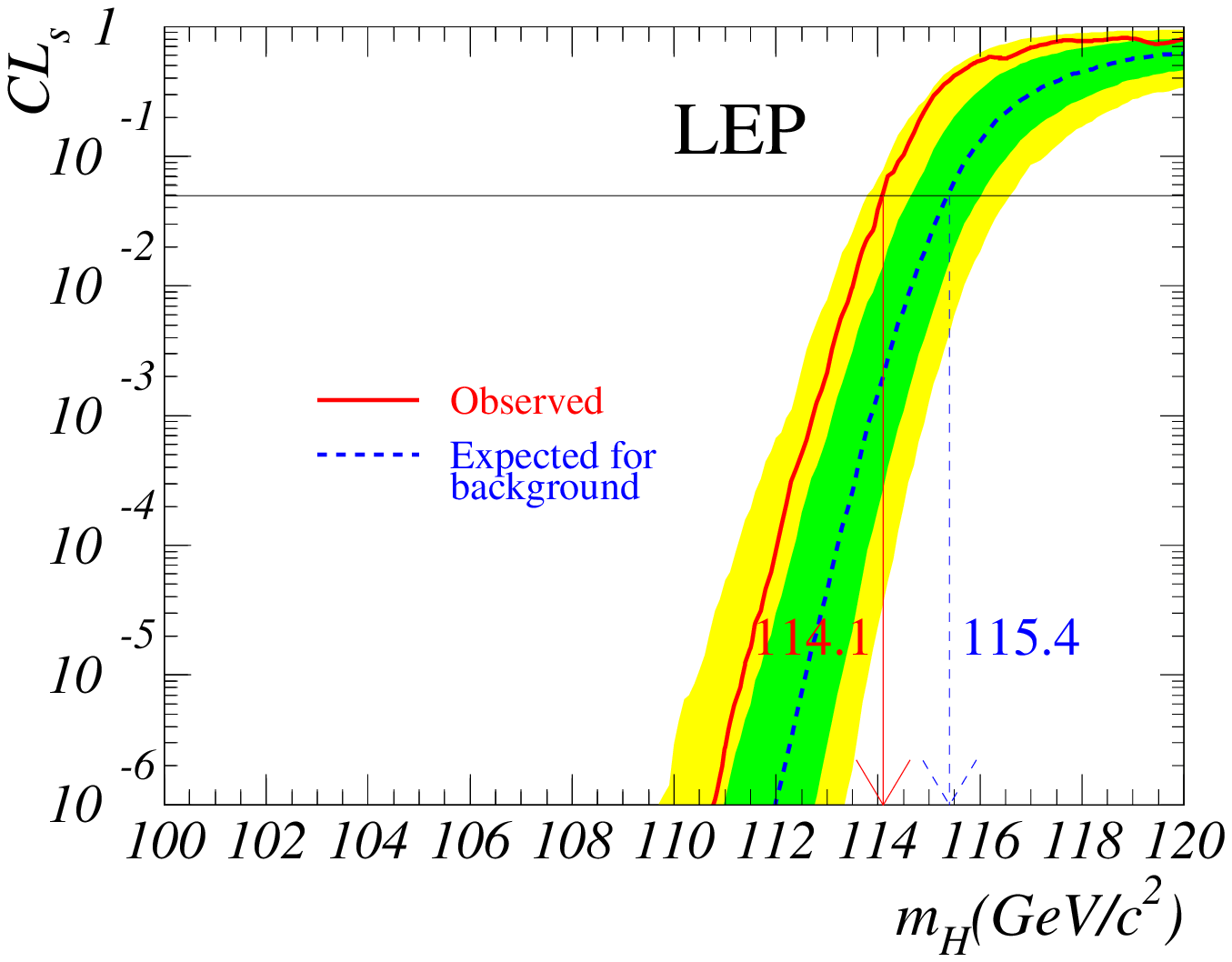,width=0.48\textwidth}
\caption[]{
Confidence levels $1-CL_{\rm b}$ and $CL_{\rm s}$ from combining the data collected 
by the four LEP experiments at energies from 189 to 209~GeV.
The solid curve is the observed result.
The shaded areas represent the symmetric $1\sigma$ and $2\sigma$ 
probability bands. 
The horizontal line at $CL_{\rm s}=0.05$ gives the mass limits at 95\% CL.
\label{sm-cl}}
\end{center}
\end{figure}

\begin{figure}[htb]
\vspace*{-0.5cm}
\begin{center}
\begin{minipage}{0.48\textwidth}
\epsfig{figure=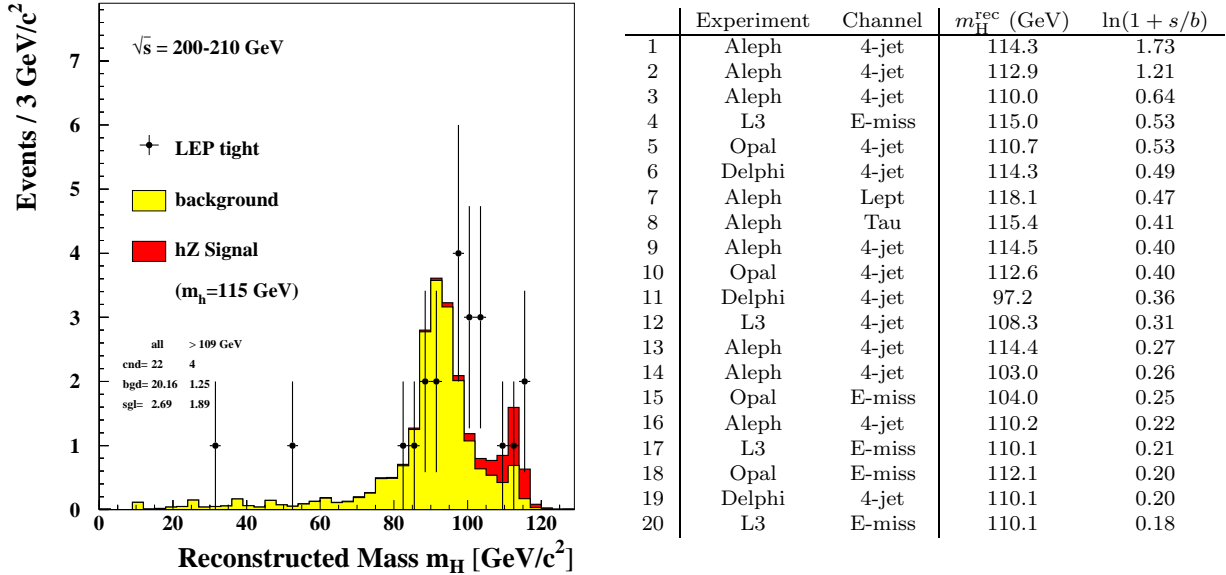,width=1.0\textwidth}
\end{minipage}
\begin{minipage}{0.48\textwidth}
\footnotesize
\begin{tabular}{c|cc|cc}
   & Experiment&  Channel &   $m_{\rm H}^{\rm rec}$ (GeV) &  $\ln (1+s/b)$ \\
\hline
1&  Aleph  &  4-jet &   114.3   & 1.73\\
2&  Aleph  &  4-jet &   112.9   & 1.21\\
3&  Aleph  &  4-jet &   110.0   & 0.64\\
4&  L3     &  E-miss&   115.0   & 0.53\\
5&  Opal   &  4-jet &   110.7   & 0.53\\
6&  Delphi &  4-jet &   114.3   & 0.49\\
7&  Aleph  &  Lept  &   118.1   & 0.47\\
8&  Aleph  &  Tau   &   115.4   & 0.41\\
9&  Aleph  &  4-jet &   114.5   & 0.40\\
10& Opal   &  4-jet &   112.6   & 0.40\\
11& Delphi &  4-jet &    97.2   & 0.36\\
12& L3     &  4-jet &   108.3   & 0.31\\
13& Aleph  &  4-jet &   114.4   & 0.27\\
14& Aleph  &  4-jet &   103.0   & 0.26\\
15& Opal   &  E-miss&   104.0   & 0.25\\ 
16& Aleph  &  4-jet &   110.2   & 0.22\\
17& L3     &  E-miss&   110.1   & 0.21\\
18& Opal   &  E-miss&   112.1   & 0.20\\
19& Delphi &  4-jet &   110.1   & 0.20\\ 
20& L3     &  E-miss&   110.1   & 0.18\\
\end{tabular}
\end{minipage}
\vspace*{-0.5cm}
\caption[]{
Left: Distribution of the reconstructed SM Higgs boson mass in searches 
conducted at energies between 200 and 210 GeV.
The figure displays the data
(dots with error bars), the predicted SM background
and the prediction for a Higgs boson of 115~GeV mass.
The number of data events selected with mass larger than 109 GeV
is 4, while 1.25 are expected from SM background processes and
1.89 from a 115 GeV signal.
Right: Properties of the candidates with the highest 
signal-over-background ratio $\ln (1+s/b)$ at 115~GeV.
The corresponding expected signal and background rates are 8.8 and 16.5 events, respectively.
\label{sm-mass}}
\end{center}
\end{figure}

\begin{figure}[htb]
\vspace*{-0.7cm}
\begin{center}
\epsfig{figure=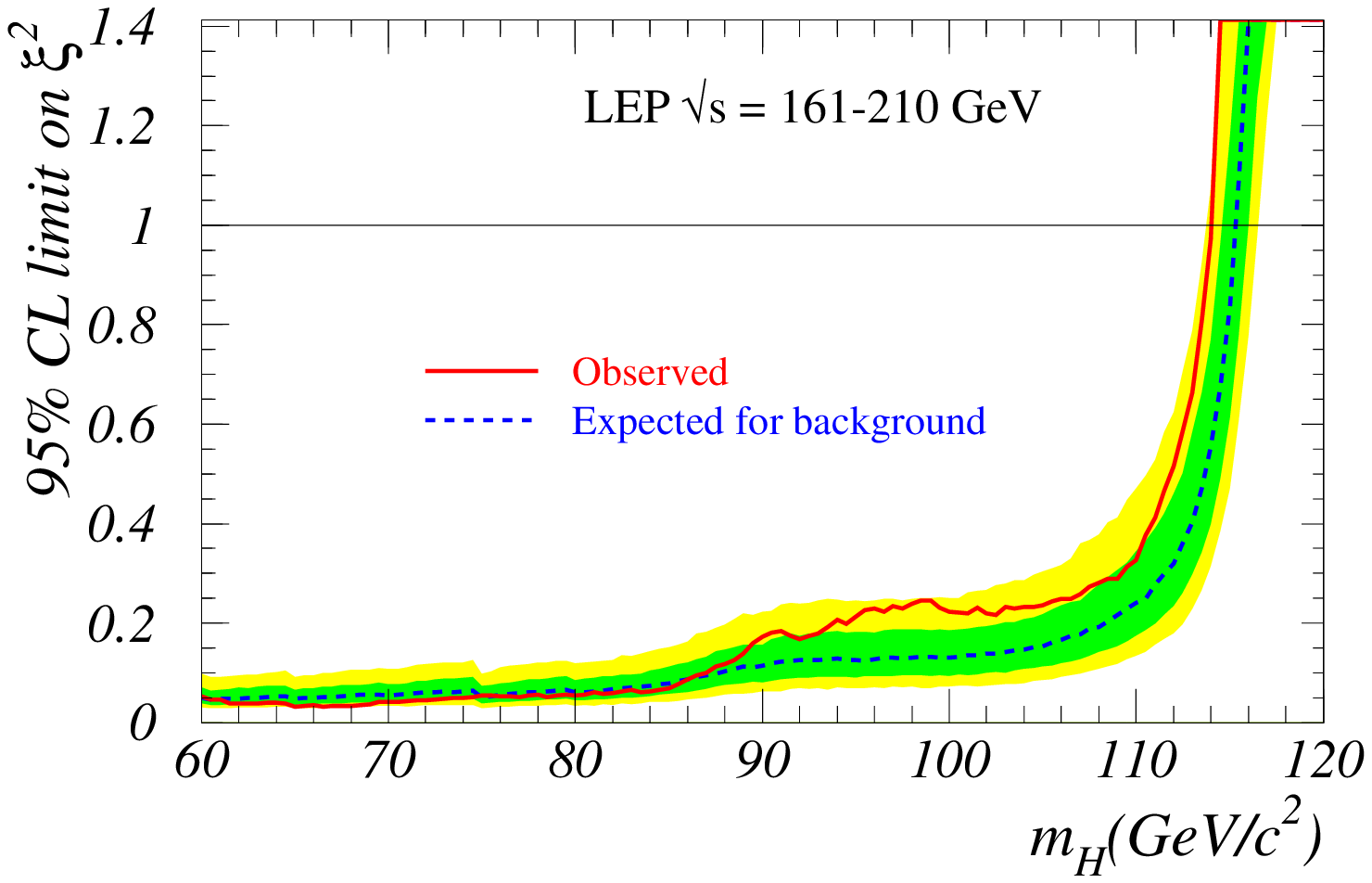,width=0.48\textwidth} 
\epsfig{figure=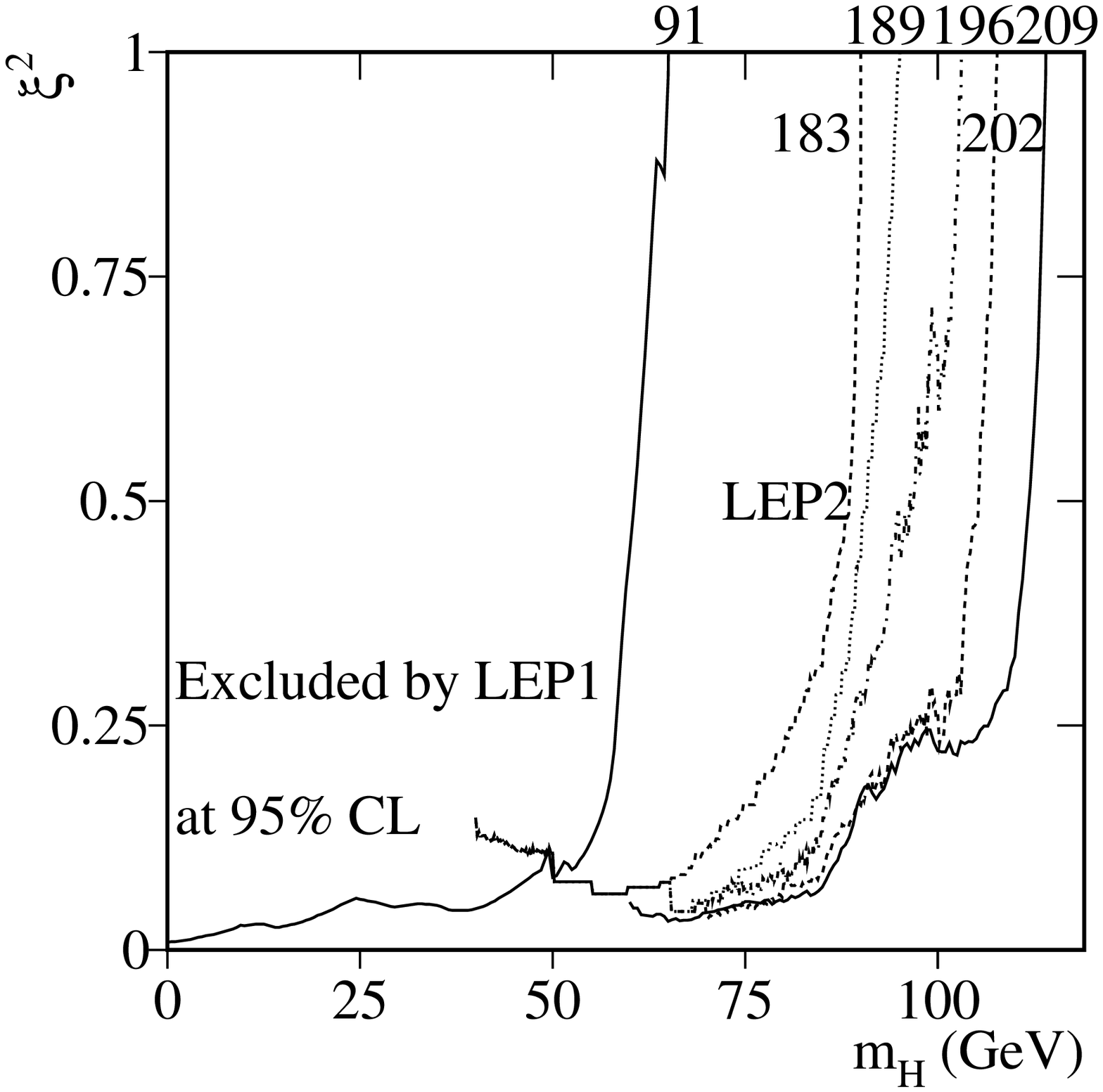,width=0.48\textwidth}
\vspace*{-0.4cm}
\caption[]{
Left:
The 95\% CL upper bound on $\xi^2$ as a function 
of \mH, where $\xi= g_{\rm HZZ}/g_{\rm HZZ}^{\rm SM}$ is the HZZ coupling
relative to the SM coupling. About $2\sigma$ deviations from the
expectation are observed at $m_{\rm H}=98$~GeV and $m_{\rm H}=115$~GeV.
Right:
The excluded $(\xi^2,m_{\rm H})$ region including 209 GeV data is compared 
with the results
from combined LEP1 data~\cite{lep1}, taken around 91 GeV center-of-mass 
energy, and previous LEP2 limits~\cite{202sall} up to 
183, 189, 196 and 202 GeV. 
The $\xi^2$ limit below 100~GeV does not become significantly stronger
when the 209 GeV data, taken in 2000, is included. 
\label{sm-xi}}
\end{center}
\end{figure}

In September 2000, ALEPH presented a data excess consistent with the
reaction 
$\rm \ee\ra HZ$ $\rm \ra \bb q\bar{q}$ for a Higgs boson mass of about 
115~GeV~\cite{alephlepc0900} which
was not confirmed by the other LEP experiments.
In November 2000, L3 provided support for a signal observation with a 
$\rm HZ\ra \bb \nu\bar{\nu}$ candidate event at the same mass.
For the summer conferences in 2001, 
all experiments updated their analyses and the probability of the 
data to be consistent with the Standard Model (SM) background
is increased,
as detailed in Table~\ref{tab:history}~\cite{sm}.
The confidence levels $CL_{\rm b}$ for a signal observation 
and $CL_{\rm s}$ for setting mass limits are shown in Fig.~\ref{sm-cl}.
The resulting SM Higgs boson mass limit is
114.1~GeV at 95\% CL.
The reconstructed mass distribution and a list of candidate events
are shown in Fig.~\ref{sm-mass}. In extensions of the SM the HZZ coupling 
might be weaker and thus the production cross section
is reduced. Figure~\ref{sm-xi} shows limits on the reduction factor
at 95\% CL.
Even if the SM cross section is reduced by a factor three,
a Higgs boson mass up to 110~GeV is excluded. 

\begin{table}[htb]
\begin{center}
\begin{tabular}{l|cccc|c|c}
              & ALEPH                & DELPHI     & L3        & OPAL    & LEP  & Significance ($\sigma$) \\
\hline
Sep. 5, 2000$^*$ & $1.6\times 10^{-4}$  & 0.67       & 0.84   & 0.47& $2.5\times 10^{-2}$ & 2.2 \\
Nov. 3, 2000 & $6.5\times 10^{-4}$&0.68&$6.8\times 10^{-2}$   & 0.19& $4.2\times 10^{-3}$ & 2.9 \\
Summer 2001 & $2.6\times 10^{-3}$&0.77&0.32&0.20&$3.4\times 10^{-2}$&2.1 
\end{tabular}
\vspace*{-0.5cm}
\caption{Background probabilities $1-CL_{\rm b}$ at a Higgs boson test-mass of 
$m_{\rm H}=115$~GeV, for the individual experiments and for the LEP data combined. 
$(^*)$ The results presented at the LEPC of September 5 were revised 
for the LEPC of November 3. The revised values are listed.
\label{tab:history}}
\end{center}
\end{table}

\clearpage
\section{MSSM Benchmark Results}

\begin{figure}[htb]
\begin{center}
\vspace*{-0.6cm}
\epsfig{figure=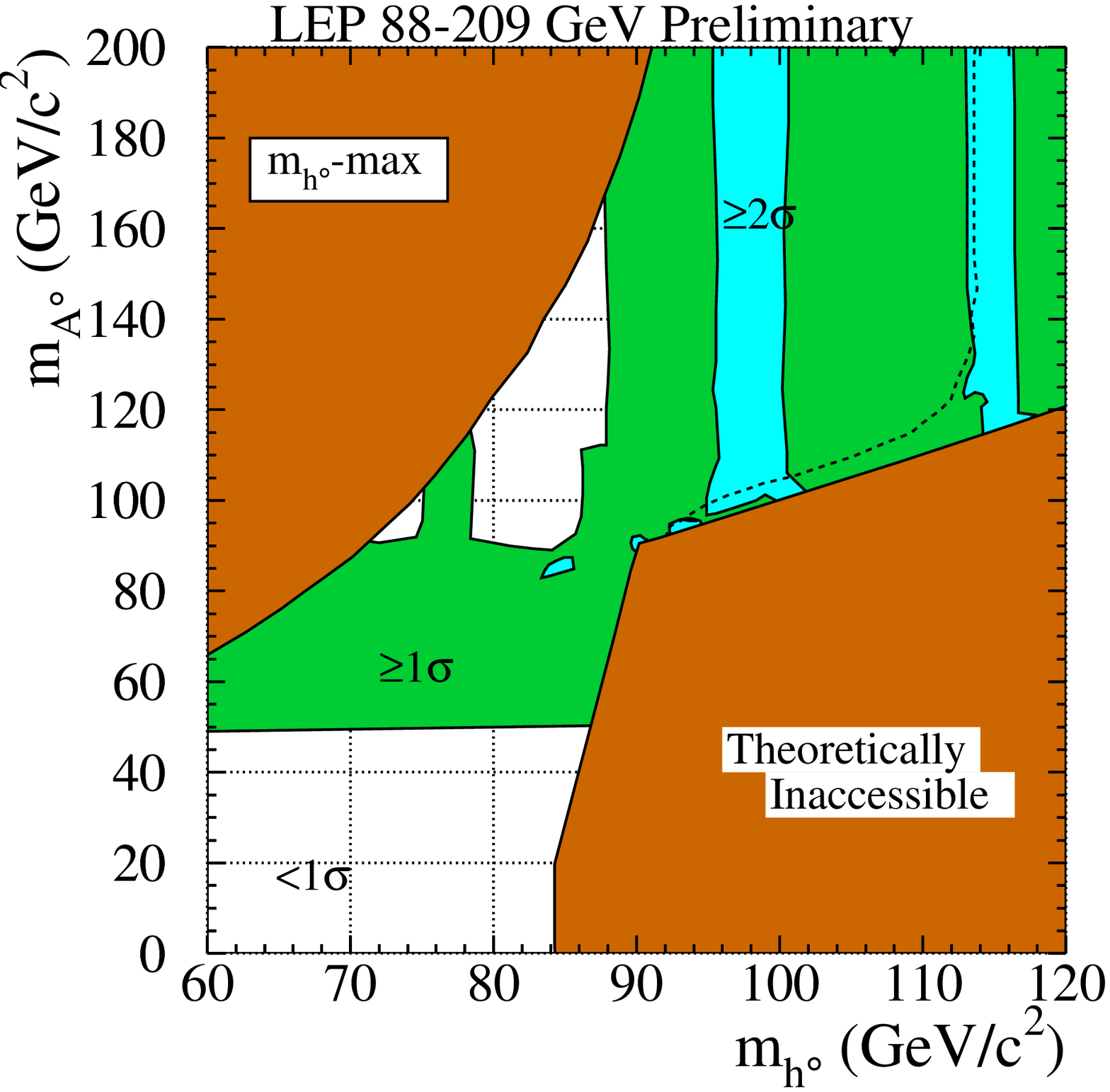,width=0.48\textwidth}
\epsfig{figure=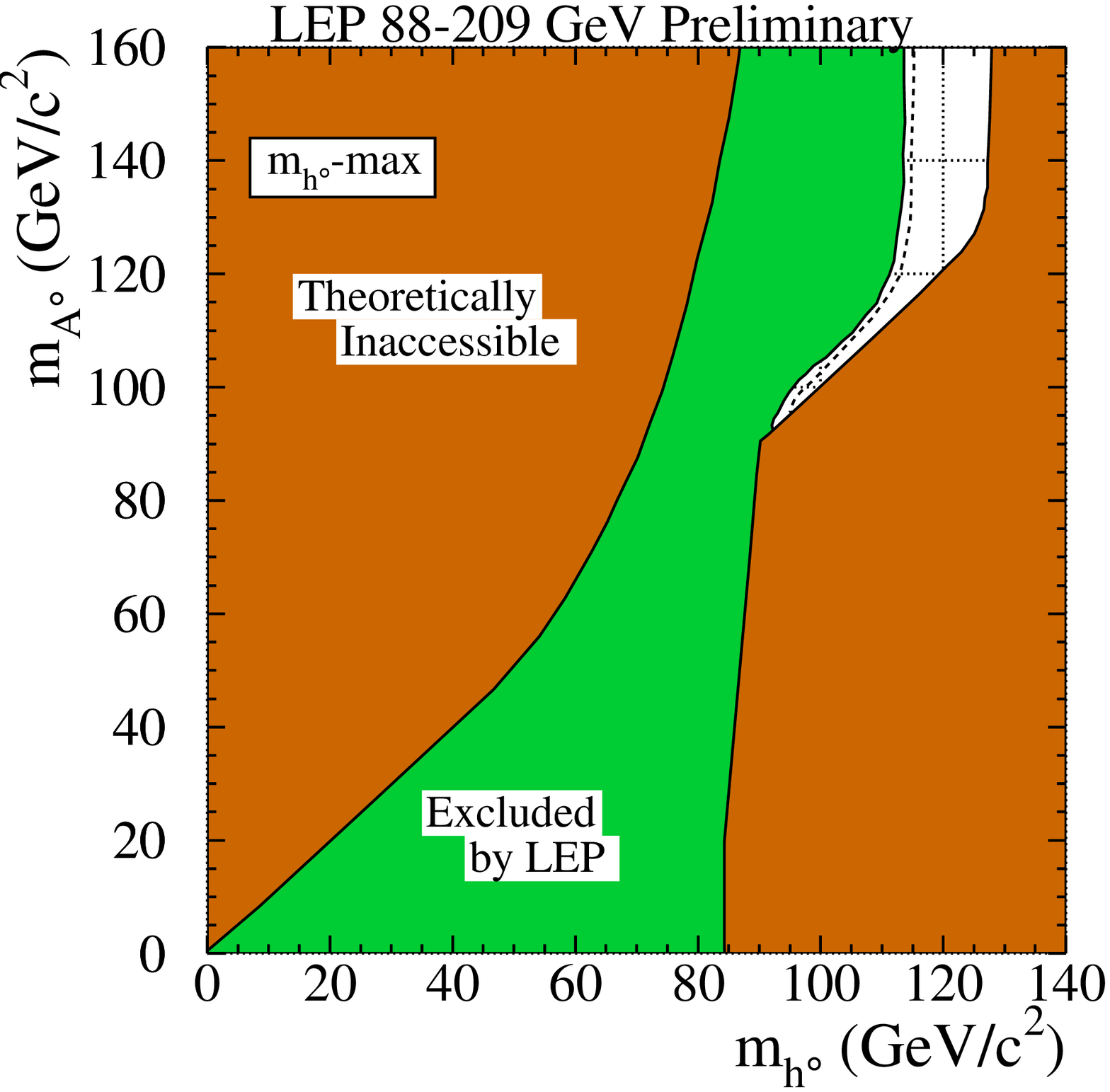,width=0.48\textwidth}
\epsfig{figure=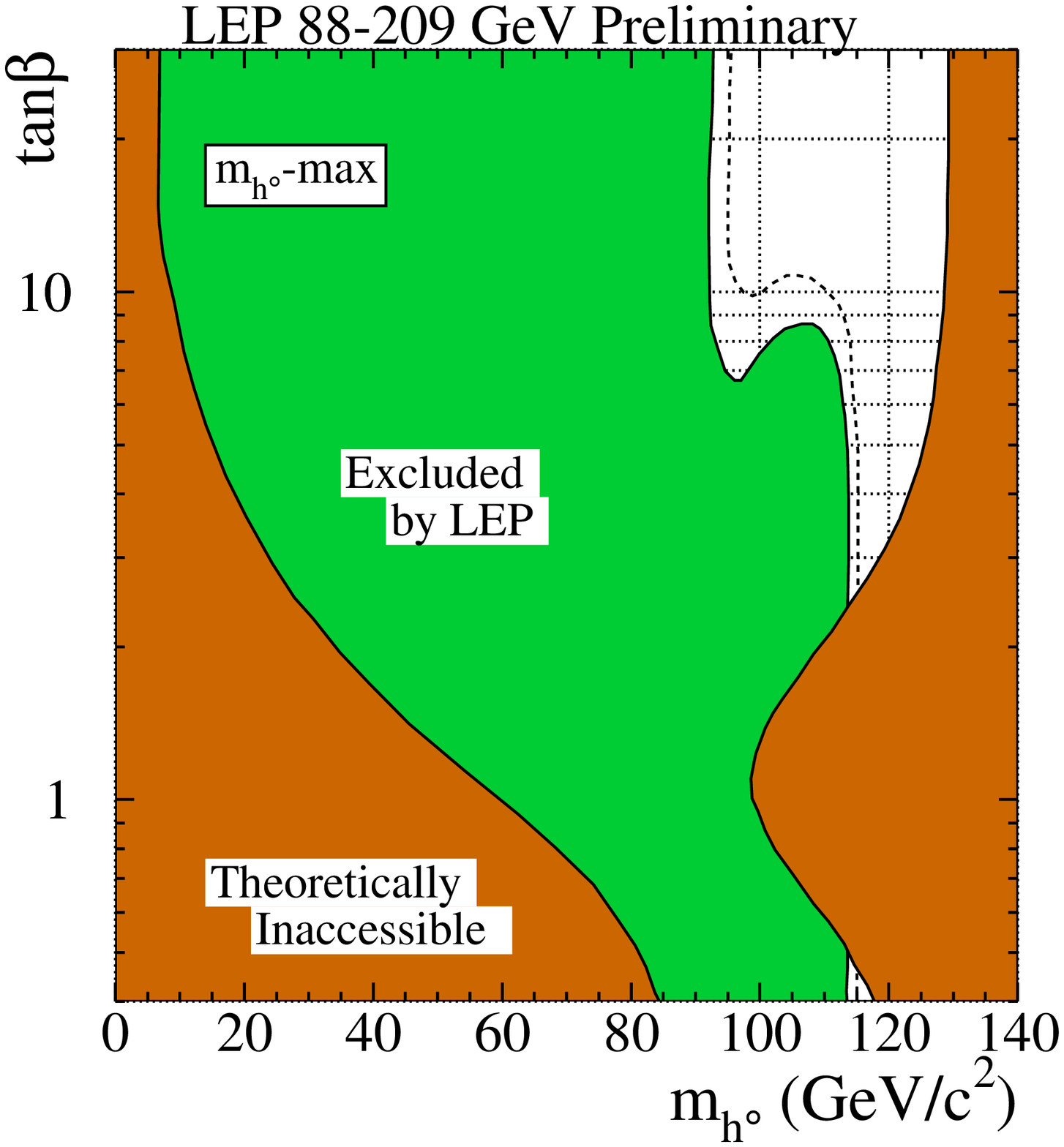,width=0.48\textwidth}
\epsfig{figure=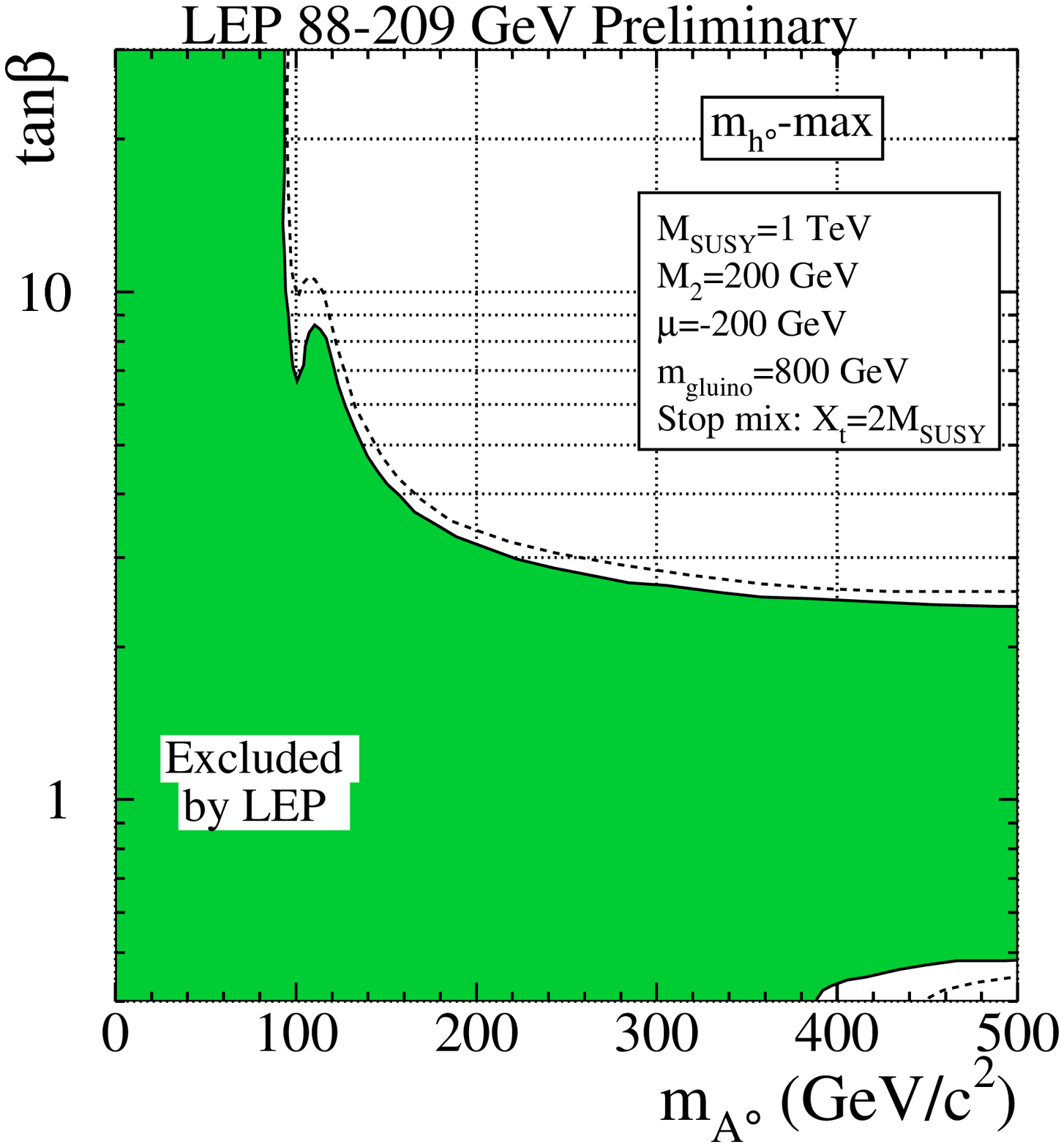,width=0.48\textwidth}
\caption[]{
Upper left:
Distribution of the discovery confidence level $1-CL_{\rm b}$ for the \mh-max 
benchmark, projected onto the ($\mh,\mA$) plane by combining the data of 
the four LEP experiments at energies from 88 to 209~GeV. 
In the white domain the observation either shows a deficit or is 
less than $1\sigma$ above the background prediction;
in the domains labelled $\ge1\sigma$ and $\ge2\sigma$ the observation 
is between 1$\sigma$ and 2$\sigma$ and 
larger than 2$\sigma$ above the prediction, respectively.
The other plots show the 95\% CL bounds on \mh, \mA\ and \tanb\ for the 
\mh-max benchmark.
The full lines represent the actual observation and
the dashed lines the limits expected on the basis of `background only' 
Monte Carlo experiments.
Upper right: projection (\mh,\mA);
lower left: projection (\mh,\tanb); 
lower right: projection (\mA,\tanb).  
\label{mssm-clb}}
\vspace*{-0.6cm}
\end{center}
\end{figure}

The Minimal Supersymmetric extension of the Standard Model (MSSM)
is the most attractive alternative to the SM.
The $\rm \ee~\ra~hA$ and $\rm \ee~\ra~hZ$ production cross sections are
complementary.
The LEP experiments have searched for the reactions
$\rm \ee~\ra~hA$ $\ra~\bb\bb$ and $\bb\tautau$.
Confidence levels $CL_{\rm b}$ and $CL_{\rm s}$ are given in 
Fig.~\ref{mssm-clb}
for the so-called benchmark results in the MSSM for large mixing 
in the scalar-top sector (\mh-max)~\cite{mssm}.

\clearpage
\section{A General MSSM Parameter Scan} 

\begin{figure}[htp]
\vspace*{-0.5cm}
\begin{center}
\begin{minipage}{0.48\textwidth}
\epsfig{figure=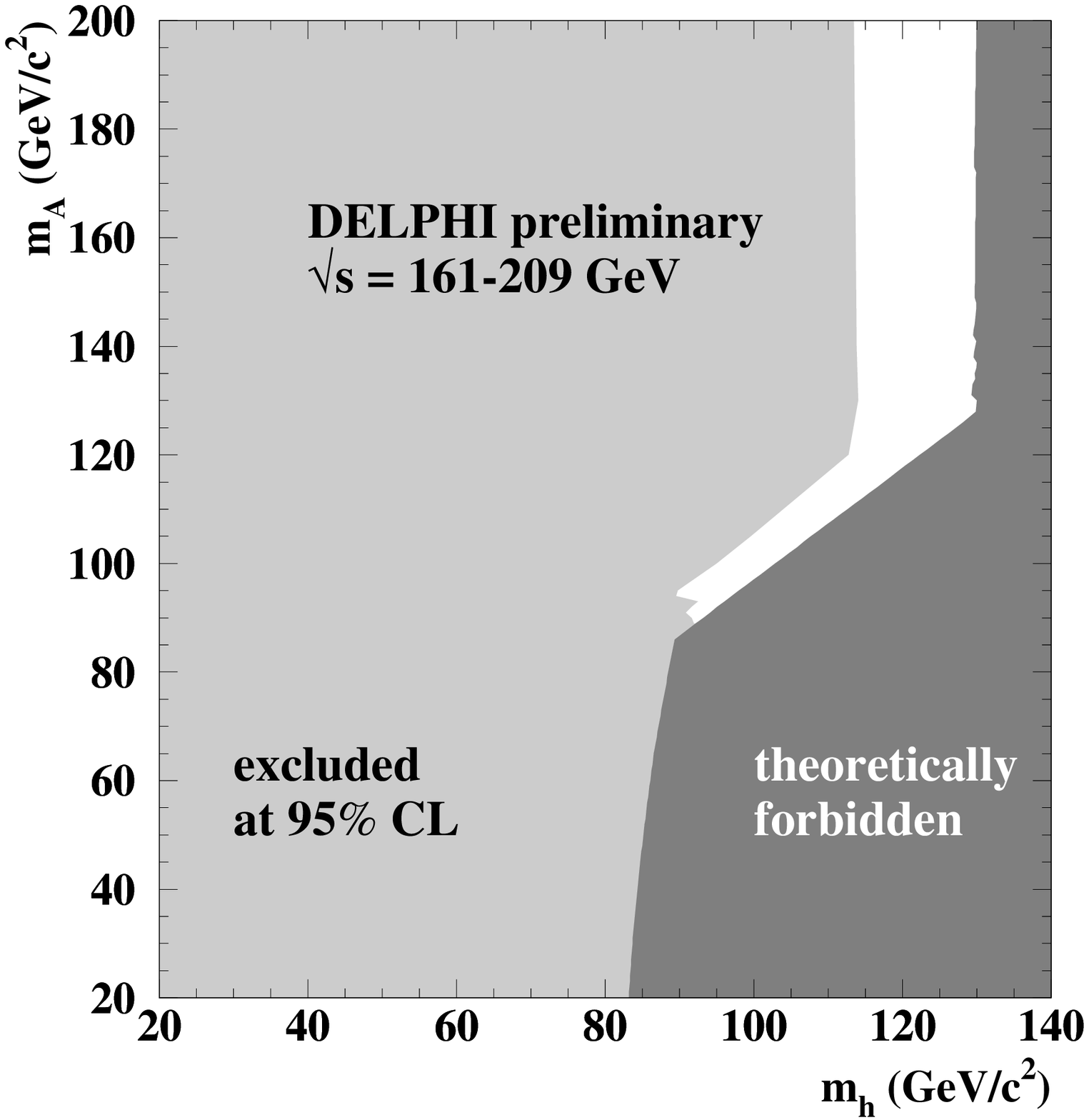,width=1.0\textwidth} 
\end{minipage}
\begin{minipage}{0.48\textwidth}
\epsfig{figure=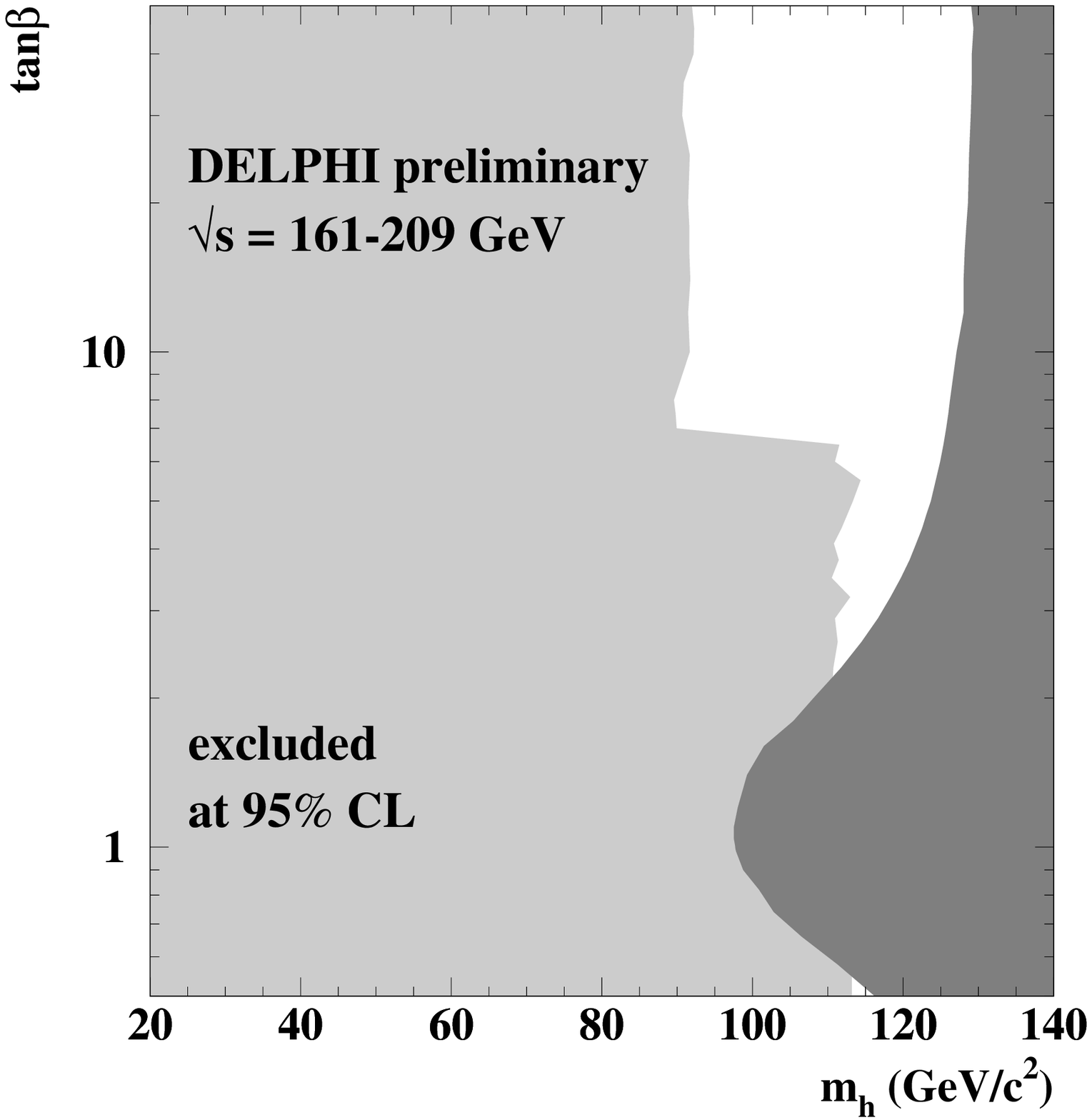,width=1.0\textwidth} 
\end{minipage}
\begin{minipage}{0.48\textwidth}
\epsfig{figure=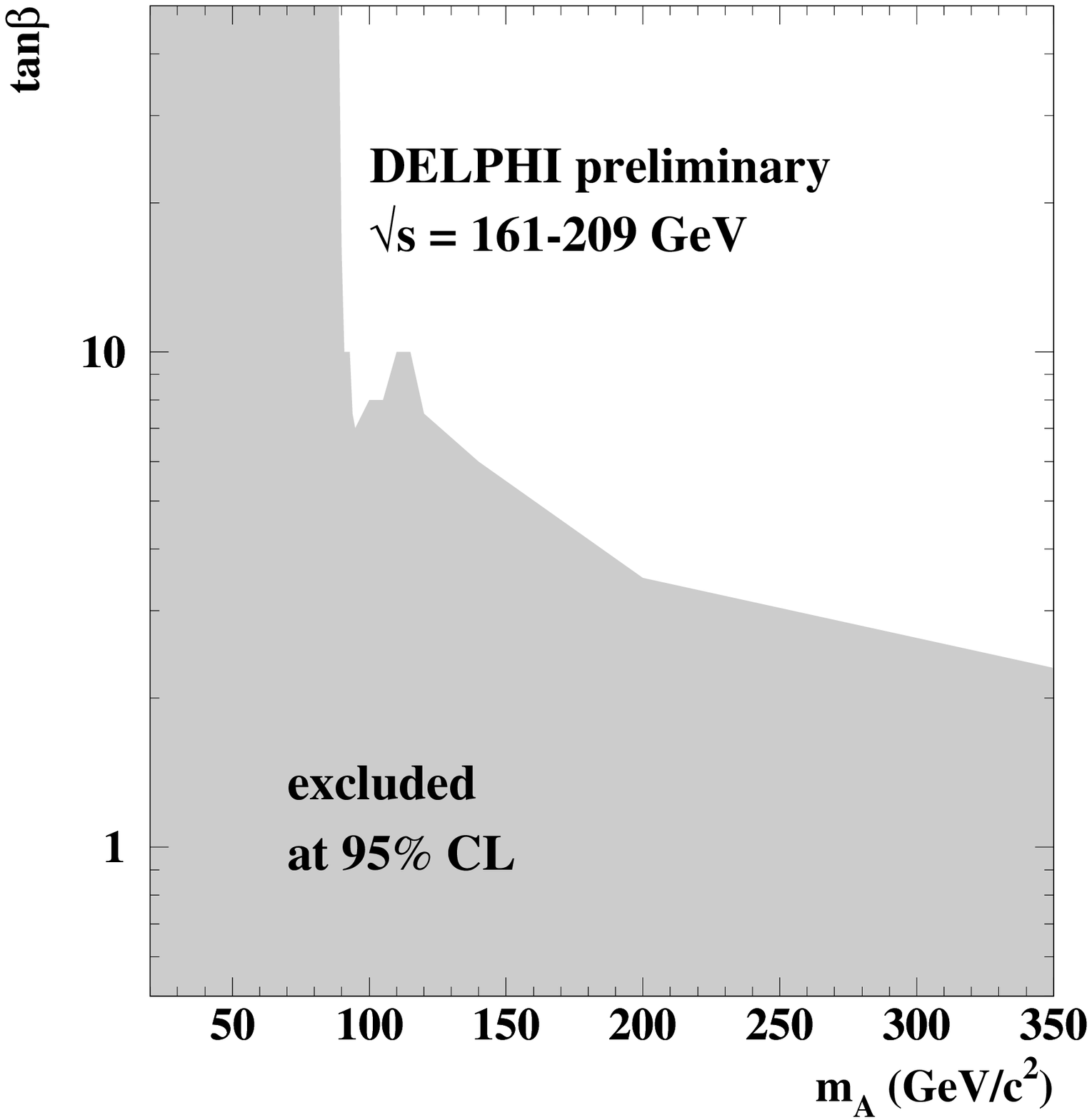,width=1.0\textwidth} 
\end{minipage}
\begin{minipage}{0.48\textwidth}
\epsfig{figure=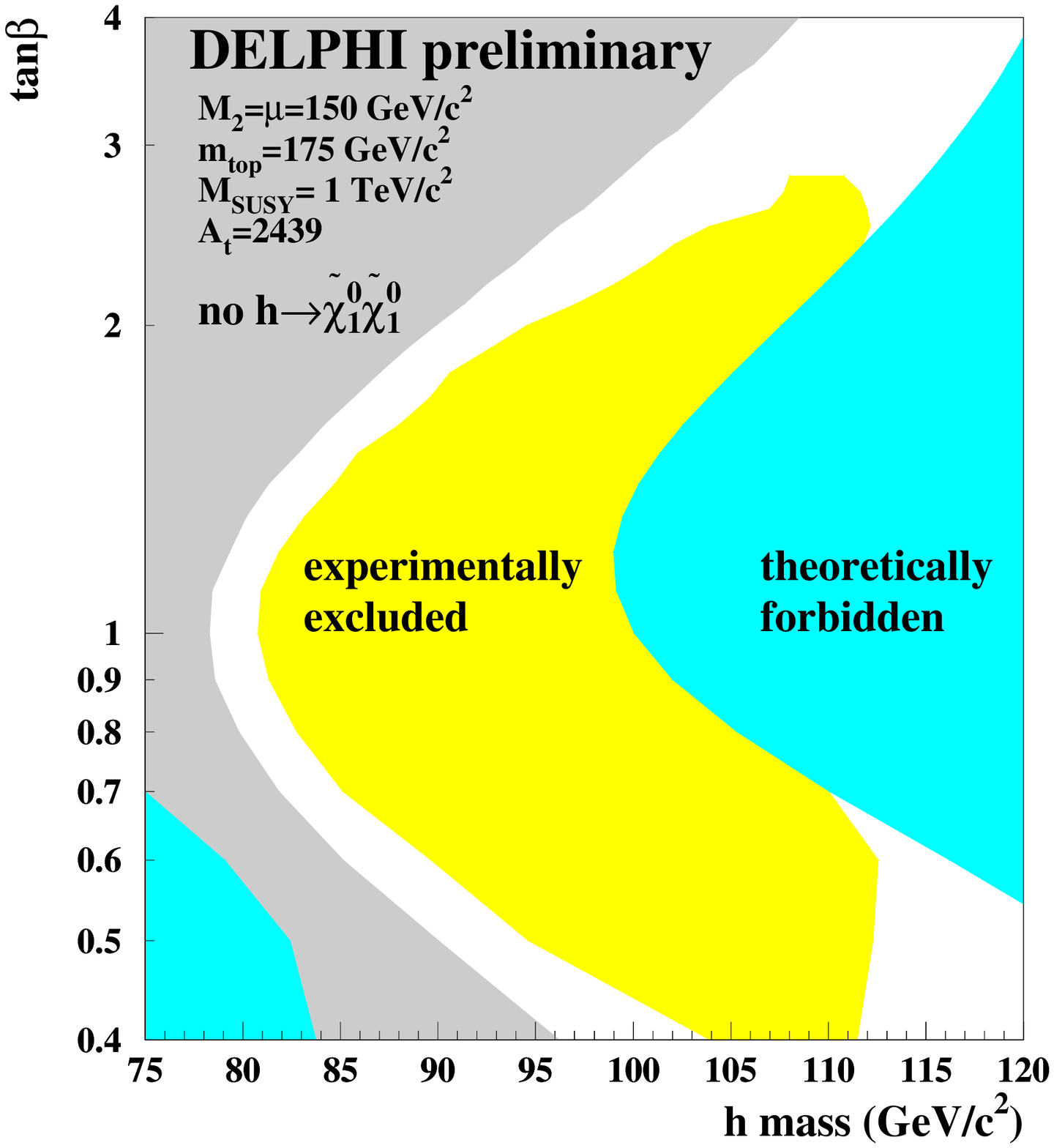,width=1.0\textwidth} 
\end{minipage}
\caption[]{
MSSM parameter scan results for DELPHI data at energies 
from 161 to 209~GeV. 
Upper left: projection (\mh,\mA);
upper right: projection (\mh,\tanb); 
lower left: projection (\mA,\tanb).  
Lower right: excluded parameter combinations
from searches for invisibly decaying Higgs bosons from 189 to 209~GeV data.
\label{mssm-scan}}
\end{center}
\vspace*{-0.4cm}
\end{figure}

Important reductions of the mass limits compared to benchmark results
were reported for LEP1 and first LEP2 data~\cite{91s,172s}.
With increasing statistics the reduction was only a few GeV by including 
the 189 GeV data of one LEP experiment (DELPHI)~\cite{189s}, and
similar for OPAL~\cite{189bsopal}.
Figure~\ref{mssm-scan} shows new results from a MSSM parameter
scan for complete DELPHI data up to 209~GeV, leading to mass limits of
89~GeV on both scalar and pseudoscalar neutral Higgs 
bosons~\cite{209s}.
These are almost identical to the DELPHI benchmark 
limits~\cite{209b}.
Figure~\ref{mssm-scan} (lower right) shows the importance of searches for 
invisible Higgs bosons~\cite{delphiinv} 
which could decay into neutralinos for some
parameter combinations of the scan (outside of the benchmark).

\begin{figure}[htb]
\begin{center}
\vspace*{-0.4cm}
\epsfig{figure=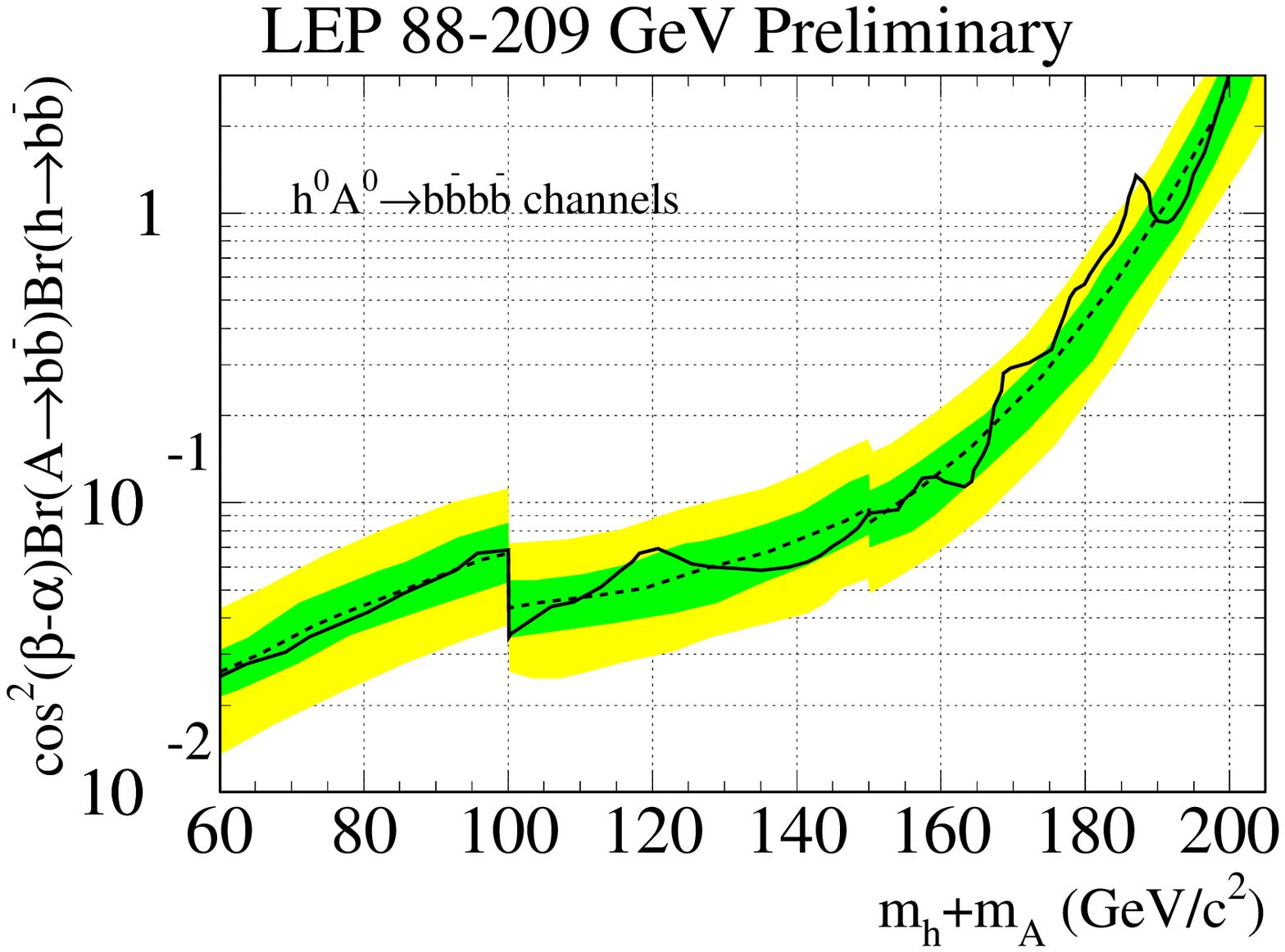,width=0.48\textwidth}
\epsfig{figure=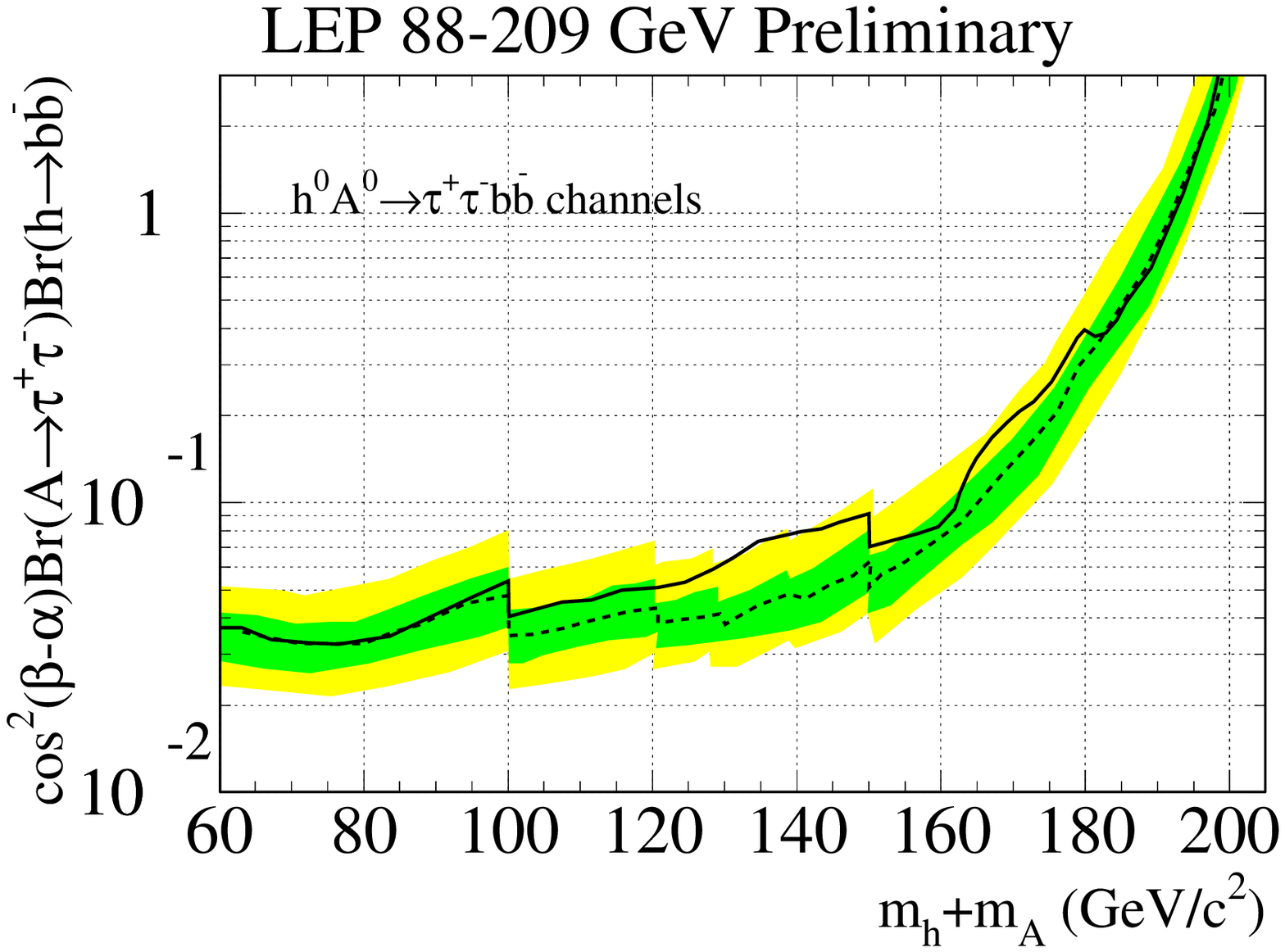,width=0.48\textwidth}
\vspace*{-0.7cm}
\caption[]{
Limits on the hA cross section as a function
of $\mh+\mA$ at 95\% CL ($\mh\approx\mA$) for the MSSM processes
\ee\ra~hA~\ra~\bb\bb\ and \bb\tautau.
This corresponds to limits of $\cos^2(\beta-\alpha)$ in the
general extension of the SM with two Higgs boson doublets.
The data of the four LEP experiments
collected at energies from 88 to 209~GeV are combined.  
The solid curve is the observed result and the dashed curve shows the 
expected median.
Shaded areas indicate the $1\sigma$ and $2\sigma$ probability bands.
\label{mssm-bbbb}}
\end{center}
\end{figure}

\begin{figure}[htb]
\begin{center}
\vspace*{-1.3cm}
\epsfig{figure=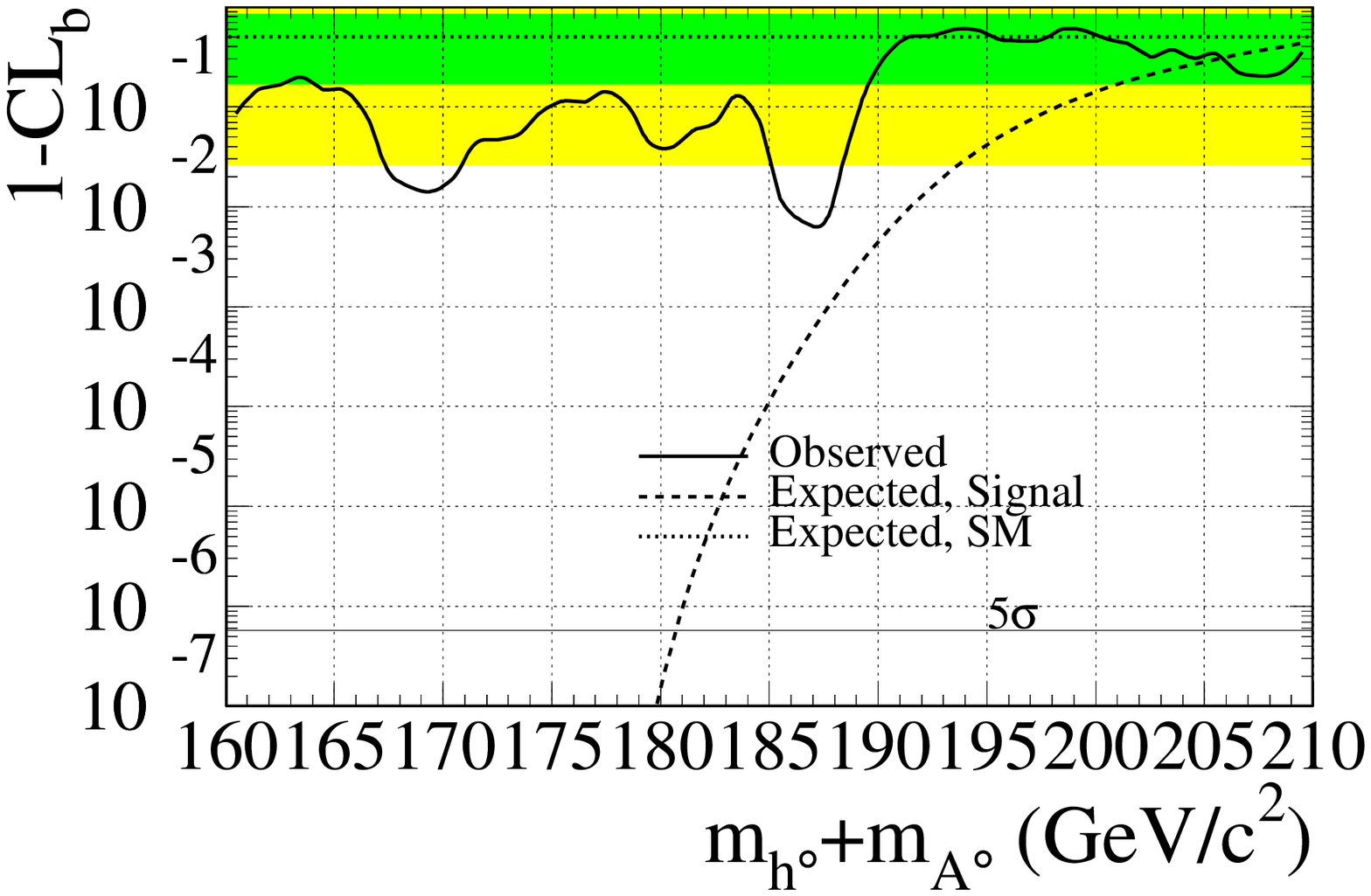,width=0.48\textwidth}
\epsfig{figure=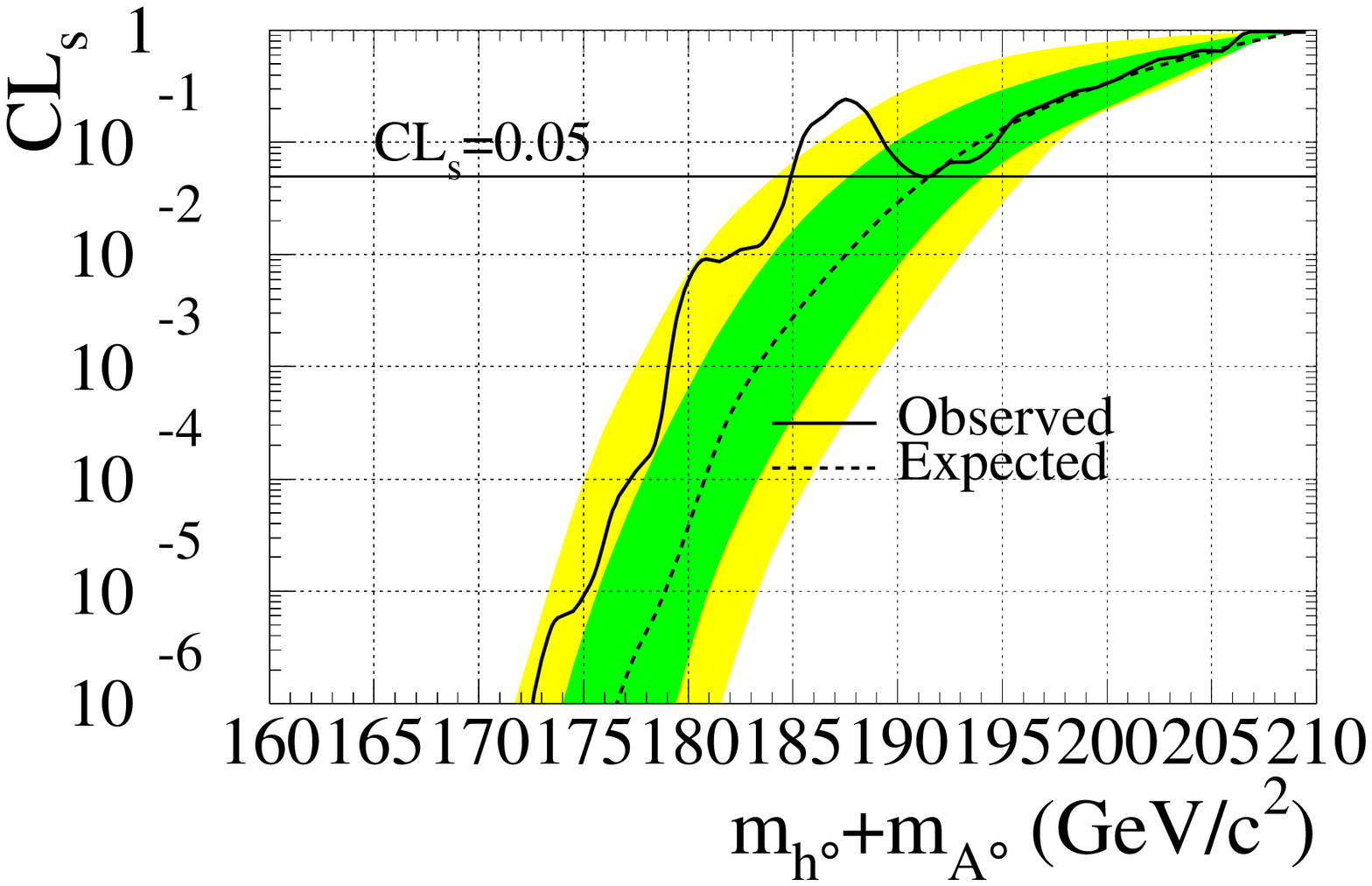,width=0.48\textwidth}
\vspace*{-0.7cm}
\caption[]{
The confidence levels $1-CL_{\rm b}$ and $CL_{\rm s}$ 
as a function of $\mh+\mA$ for the case $\mh\approx\mA$
(where only the \ee\ra~hA process contributes since $\sba\approx0$). 
The straight line at 0.5 and the shaded $1\sigma$
and $2\sigma$ probability bands represent the expected
background-only result.
The solid curve is the observed result and the dashed curve shows the 
expected median for a signal.
The horizontal line at $1-CL_{\rm b}=5.7\times 10^{-7}$ indicates the level 
for a $5\sigma$ discovery and the intersections of the curves with the 
horizontal line at $CL_{\rm s}=0.05$ give the limit on $\mh+\mA$ 
at 95\% CL.
The data of the four LEP experiments
collected at energies from 88 to 209~GeV are combined.  
\label{mssm-diag}}
\end{center}
\vspace*{-0.4cm}
\end{figure}

\clearpage
Limits on the $\rm \ee~\ra~hA~\ra~\bb\bb$ and $\bb\tautau$
production rates are given in Fig.~\ref{mssm-bbbb}~\cite{mssm}
for the example of $\mh\approx\mA$.
As noted previously for 202 GeV data, taken in 1999, 
and first 2000 data,
h and A mass limits were 2 GeV below
expectation~\cite{as2000} and this tendency is enhanced by including 209 GeV
data, in which case they are 3.1 to 3.6 GeV below the expectations of about 
95~GeV.
A possible explanation is given that the HZ excess at about 115~GeV is due 
to the heavier scalar and that, in addition, the production of hA with 
masses between 90 and 100 GeV occurs~\cite{as2000}.
Figure~\ref{mssm-bbbb} shows a data excess above $2\sigma$ 
for $m_{\rm h}+m_{\rm A}=187$~GeV in the \bb\bb\ channel.
The same data excess is also expressed in the $CL_{\rm b}$ and 
$CL_{\rm s}$ distributions as shown in Fig.~\ref{mssm-diag}.
The hypothesis of the production of three MSSM Higgs bosons is supported
by the data excess seen in Fig.~\ref{sm-mass} at 100 GeV 
which could result from hZ production in addition to HZ production.
For the reported MSSM parameters~\cite{as2000}
$\cos^2(\beta-\alpha)\approx 0.9$; therefore 
$\sin^2(\beta-\alpha)=\xi^2\approx0.1$.
The $\xi^2$ limit in the 100 GeV mass region shows a deviation of about
2$\sigma$ between expected and observed limit,
as seen in Fig.~\ref{sm-xi} (left).
Figure~\ref{sm-xi} (right) shows that this new support is only observed
in the complete LEP data.

\clearpage
\section{Charged Higgs Bosons}
\begin{figure}[htb]
\vspace*{-0.3cm}
\begin{center}
\epsfig{figure=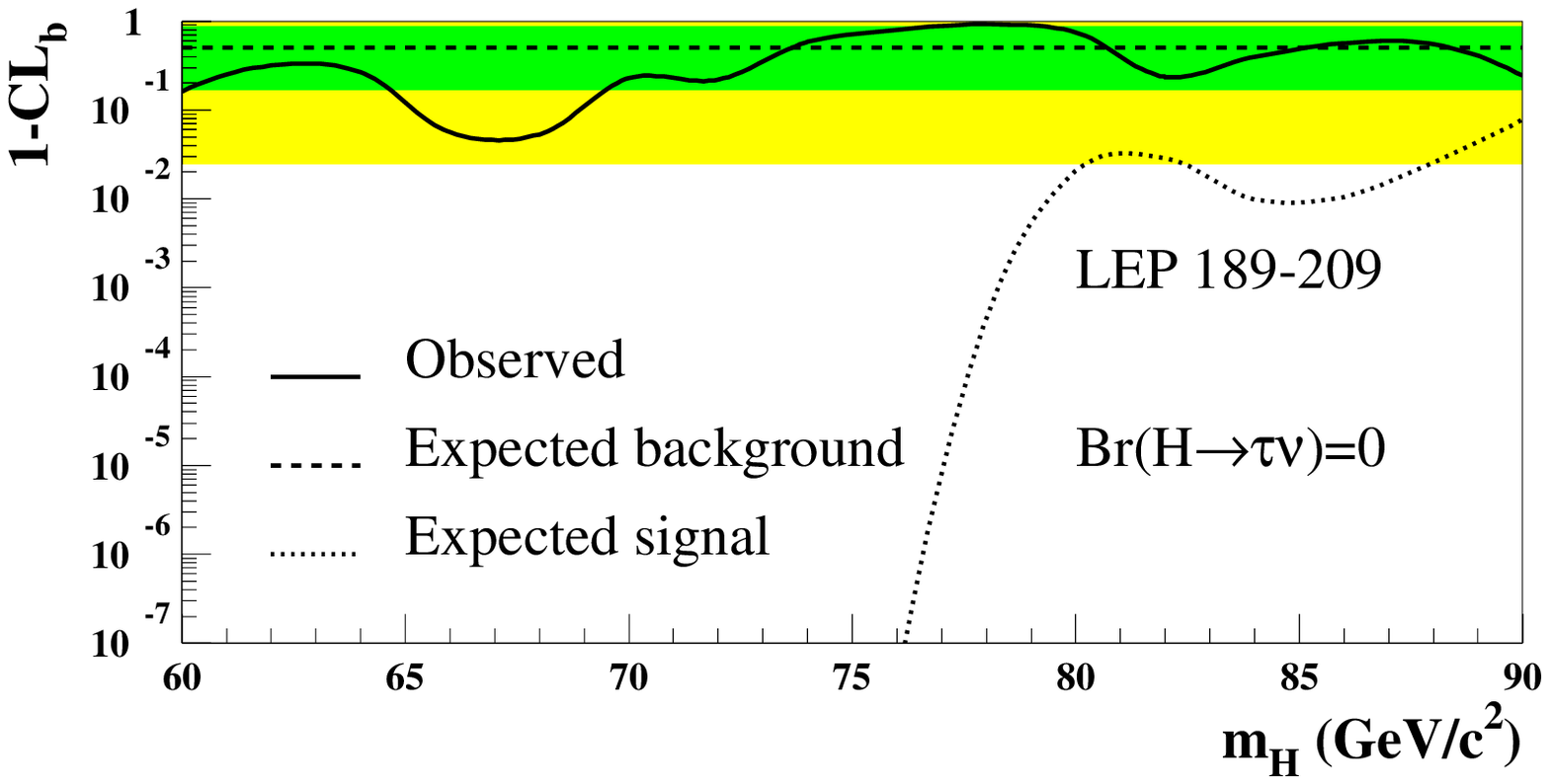,width=0.48\textwidth}
\epsfig{figure=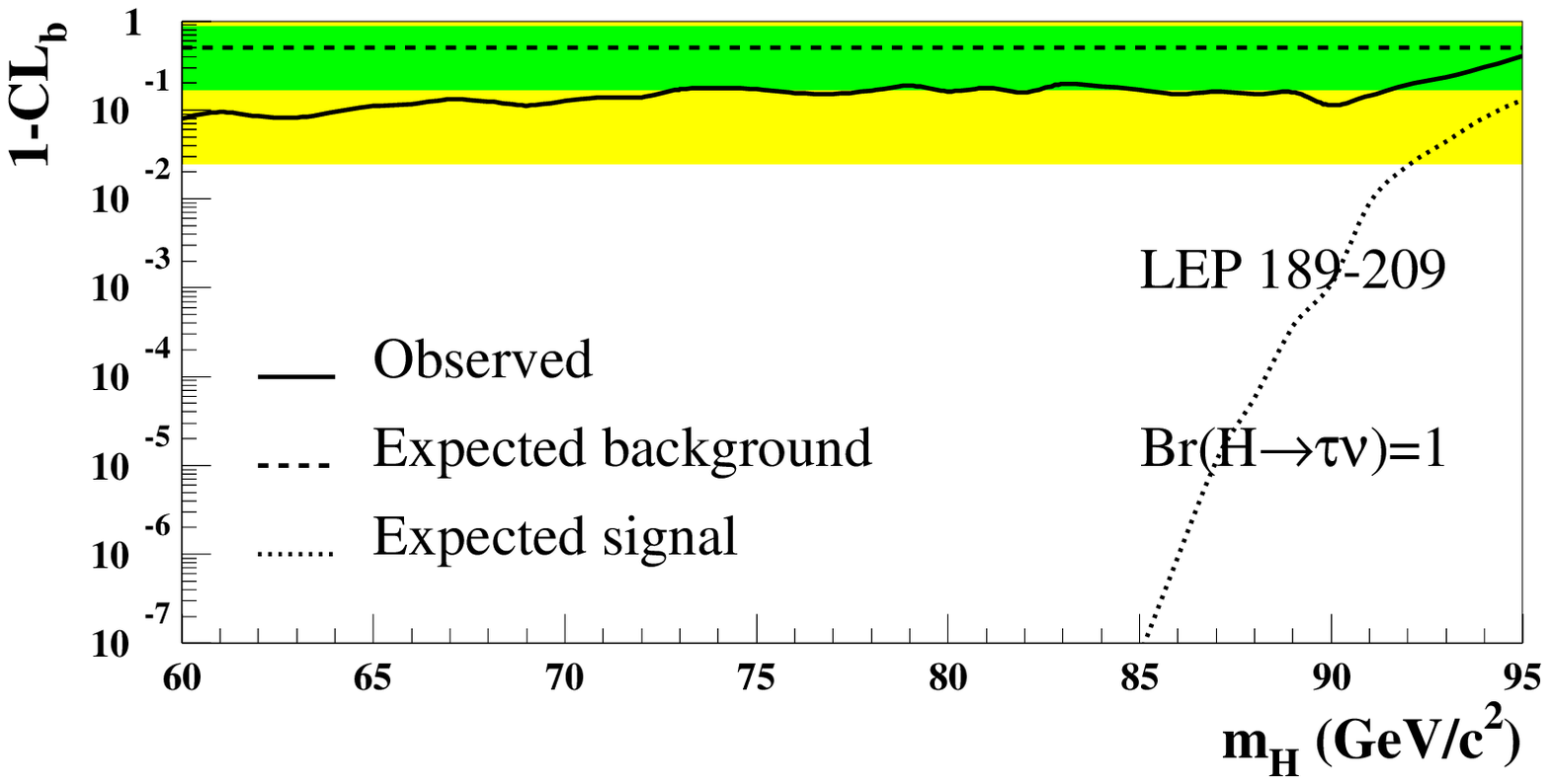,width=0.48\textwidth}
\vspace*{-0.5cm}
\caption[]{
The $1-CL_{\rm b}$ distributions for the 
$\csbar\cbars$ and $\tp\nu\tm\nubar$
channels, combining the data collected by the four LEP 
experiments at energies from 189 to 209~GeV. 
The shaded areas represent the $1\sigma$ and $2\sigma$ 
probability bands. Owing to the large irreducible $\rm WW\ra\csbar\cbars$ 
background, less sensitivity is obtained near the W mass
for the case Br(\Hp~\ra~\tp$\nu)=0$ as it can be seen from the expected
signal probability curve.
\label{charged-clb}}
\end{center}
\end{figure}

\begin{figure}[htb]
\vspace*{-0.6cm}
\begin{center}
\epsfig{figure=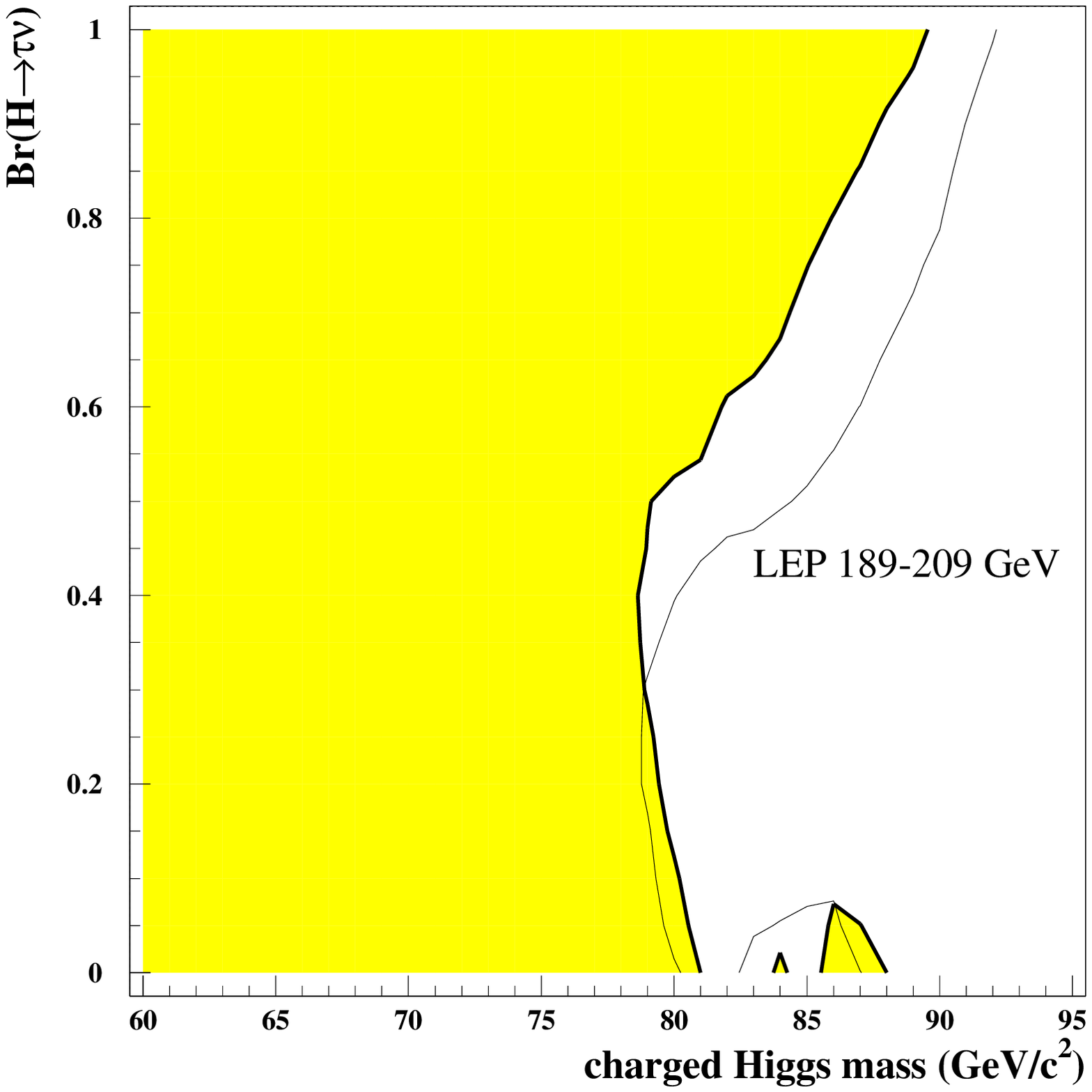,width=0.48\textwidth}
\epsfig{figure=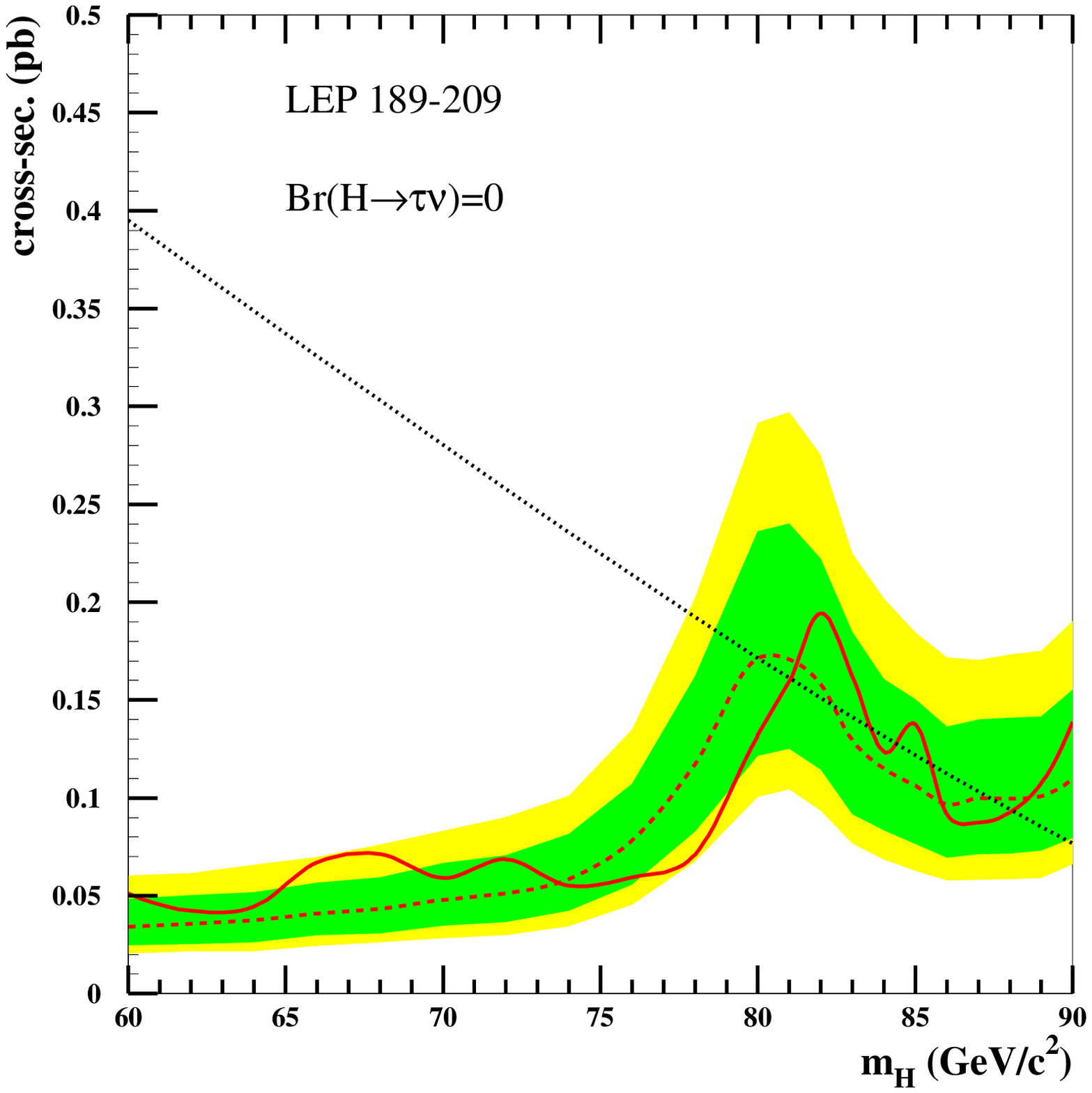,width=0.48\textwidth}
\vspace*{-0.5cm}
\caption[]{
Left:
The 95\% CL bounds on \mHpm\ as a function of the branching ratio
Br(\Hp~\ra~\tp$\nu$), combining the data collected by the four LEP 
experiments at energies from 189 to 209~GeV. 
The expected median limits are indicated by the thin line and the
observed limits by the thick line.
Right:
The 95\% CL bounds on the production cross section 
for Br(\Hp~\ra~\tp$\nu)=0$.
The observed limit is given by the solid line and the expected limits 
are indicated by the dashed line and the shaded bands 
($1\sigma$ and $2\sigma$).
The dotted line gives the expected cross section at 206 GeV center-of-mass
energy. Radiative corrections vary the cross section by typically $\pm 10$\% 
depending on the model parameters~\cite{rad1,rad2}.
\label{charged-limit}}
\end{center}
\vspace*{-0.3cm}
\end{figure}

The search for charged Higgs bosons is performed in
the framework of the general extension of the SM with
two Higgs boson doublets. The combined $CL_{\rm b}$ distributions 
from the four LEP experiments for the reactions
$\ee~\ra~\Hp\Hm~\ra~\csbar\cbars$ and $\tp\nu\tm\nubar$
are presented in Fig.~\ref{charged-clb}~\cite{hphm}.
The resulting mass limit, which also includes the cs$\tau\nu$ channel, 
is 78.6~GeV at 95\% CL and it is valid for any 
branching ratio Br(\Hp~\ra~\tp$\nu$) as shown in Fig.~\ref{charged-limit}.
The cross section limit for the $\csbar\cbars$ channel shows 
that the barrier from irreducible WW background events is almost passed.
Optimization of the analyses of each individual experiment for 
higher masses could increase the decay-mode-independent reach by more
than 5~GeV. Already a pre-LEP2 study~\cite{prelep2} based on a luminosity
of 500~pb$^{-1}$ pointed out that
`..., the kinematic region above $m_{\rm W}$ will be a challenge for the
decay-mode-independent sensitivity, it might even be unfeasible ...'.

\clearpage
\section{Invisible Higgs Boson Decays}
\begin{figure}[htb]
\begin{center}
\vspace*{-0.3cm}
\epsfig{figure=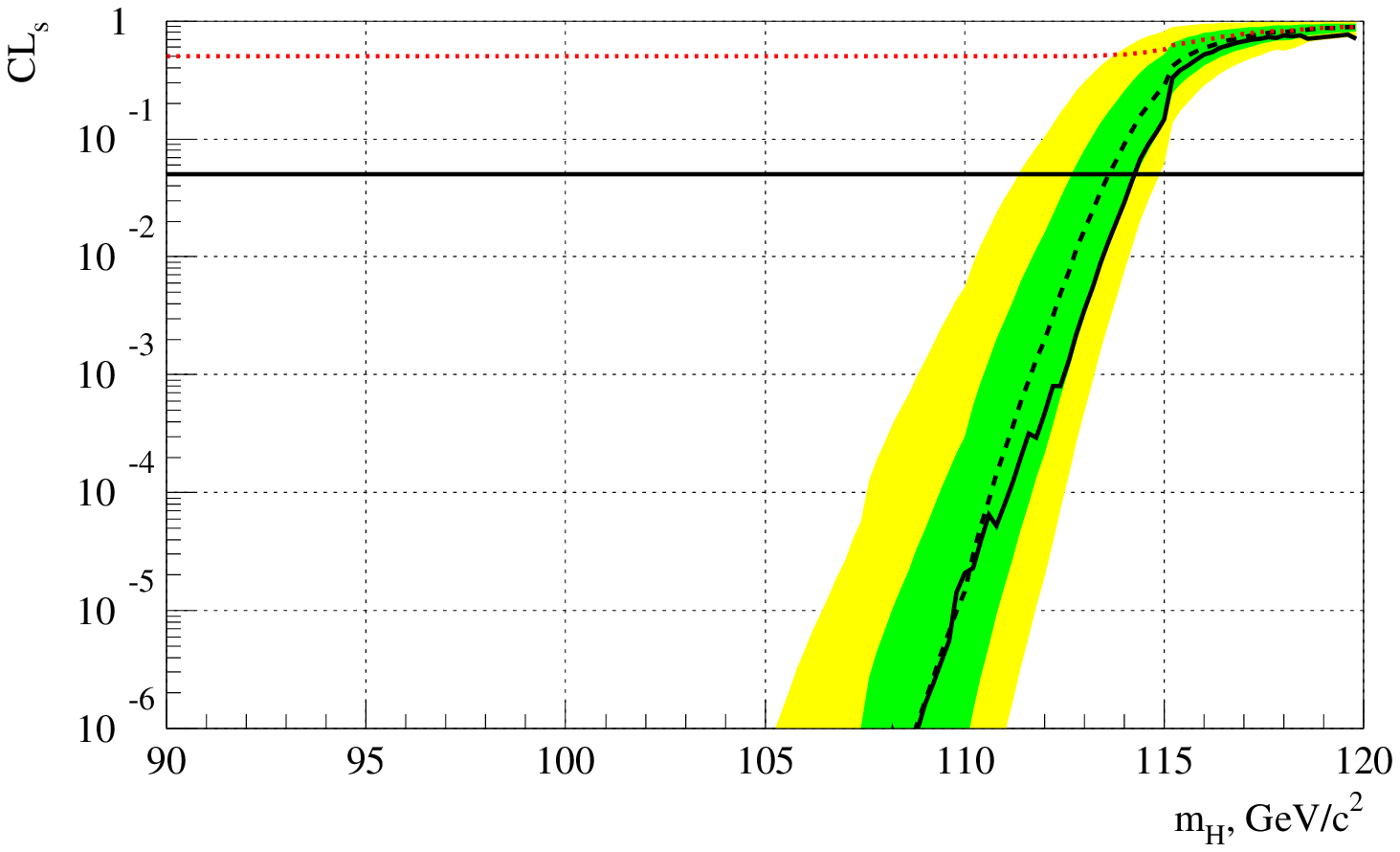,width=0.48\textwidth}
\epsfig{figure=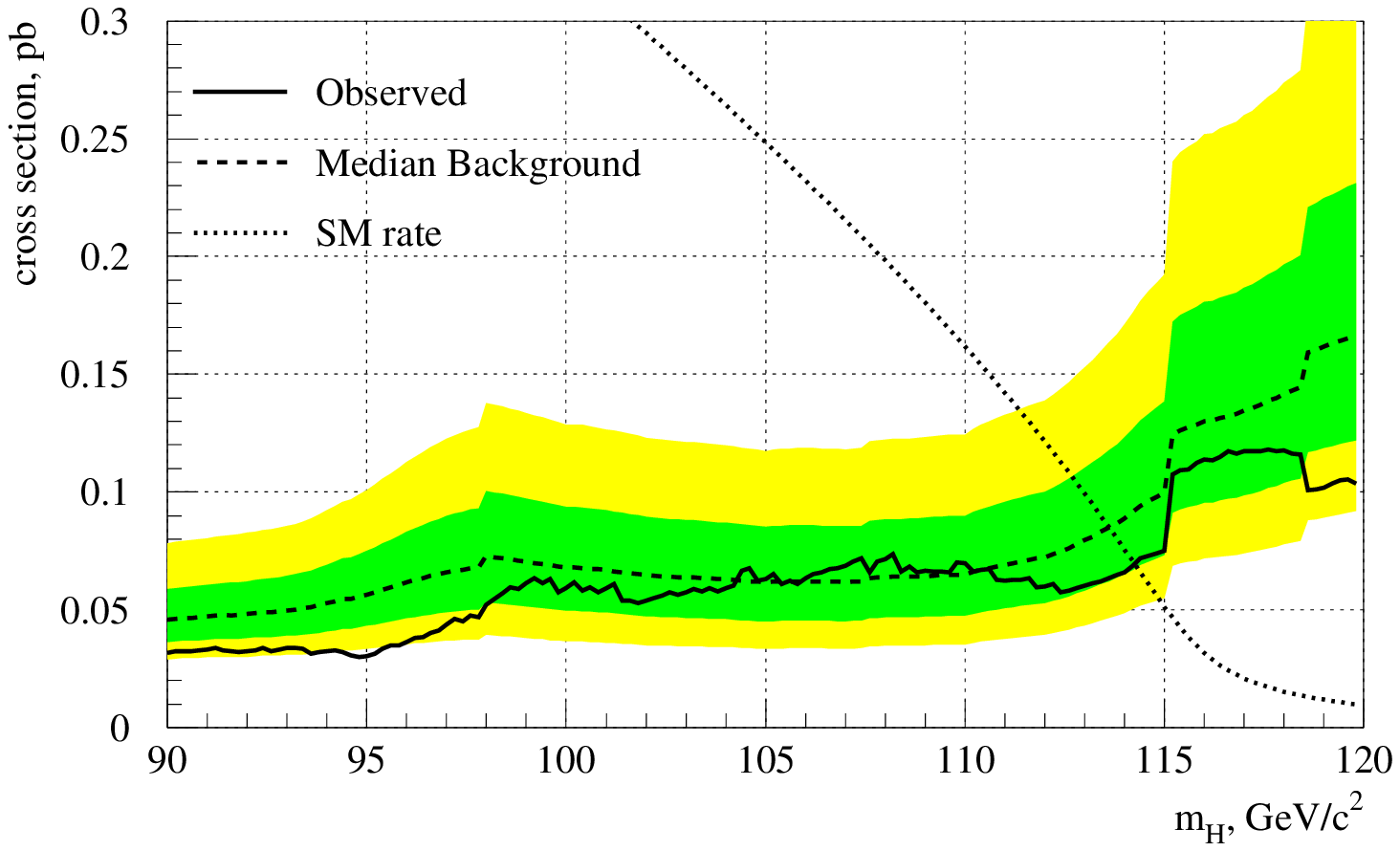,width=0.48\textwidth}
\vspace*{-0.7cm}
\caption[]{
Left:
The $CL_{\rm s}$ distribution for all hadronic and charged leptonic
channels for invisible Higgs boson decays, 
combining the data collected by the four LEP experiments 
at energies from 200 to 209~GeV.
The observed limit is given by the solid line and the expected limits 
are indicated by the dashed line and the shaded bands
($1\sigma$ and $2\sigma$).
The horizontal line at $CL_{\rm s}=0.05$ gives the limit at 95\% CL.
Right:
The 95\% CL bounds on the production cross section
for Br$(\rm H^\prime \ra invisible)=1$.
The observed limit is given by the solid line and the expected limits 
are indicated by the dashed line and the shaded bands 
($1\sigma$ and $2\sigma$).
The dotted line gives the expected cross section at 206 GeV
for Higgs bosons decaying 100\% invisibly.
\label{inv-limit}}
\end{center}
\vspace*{-0.8cm}
\end{figure}

In some models the Higgs boson can decay into invisible particles,
such as neutralinos or majorons.
The search for these Higgs bosons is performed in the Higgs 
bremsstrahlung process in association with a Z boson.
All hadronic and charged leptonic Z decays are investigated~\cite{inv}.
The $CL_{\rm s}$ distribution and the cross section limit are given
in Fig.~\ref{inv-limit}. The mass limit is 114.4~GeV.

\section{Flavour-Independent Higgs Boson Decays}
\begin{figure}[htb]
\vspace*{-0.6cm}
\begin{center}
\vspace*{-0.3cm}
\epsfig{figure=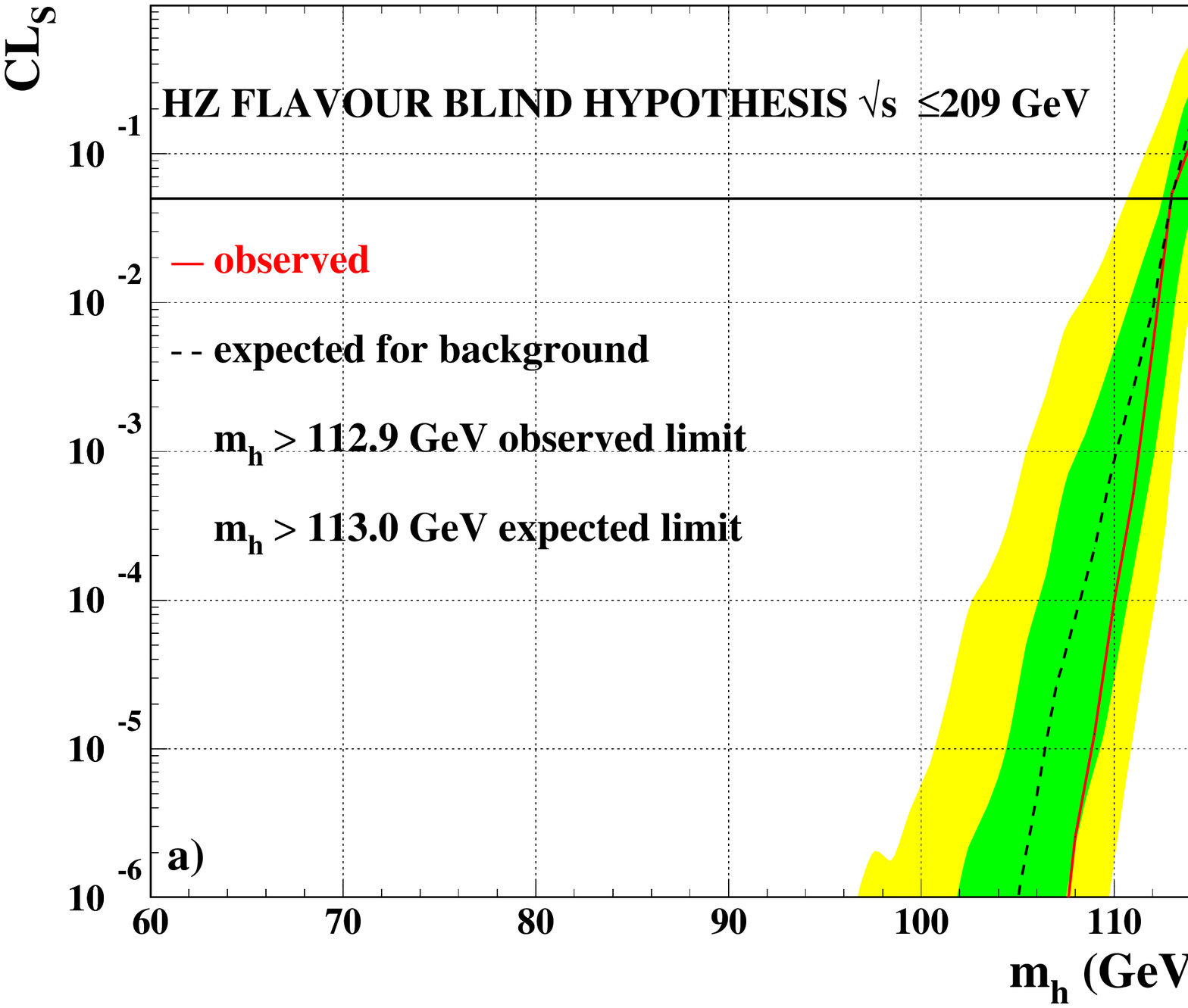,width=0.48\textwidth}
\epsfig{figure=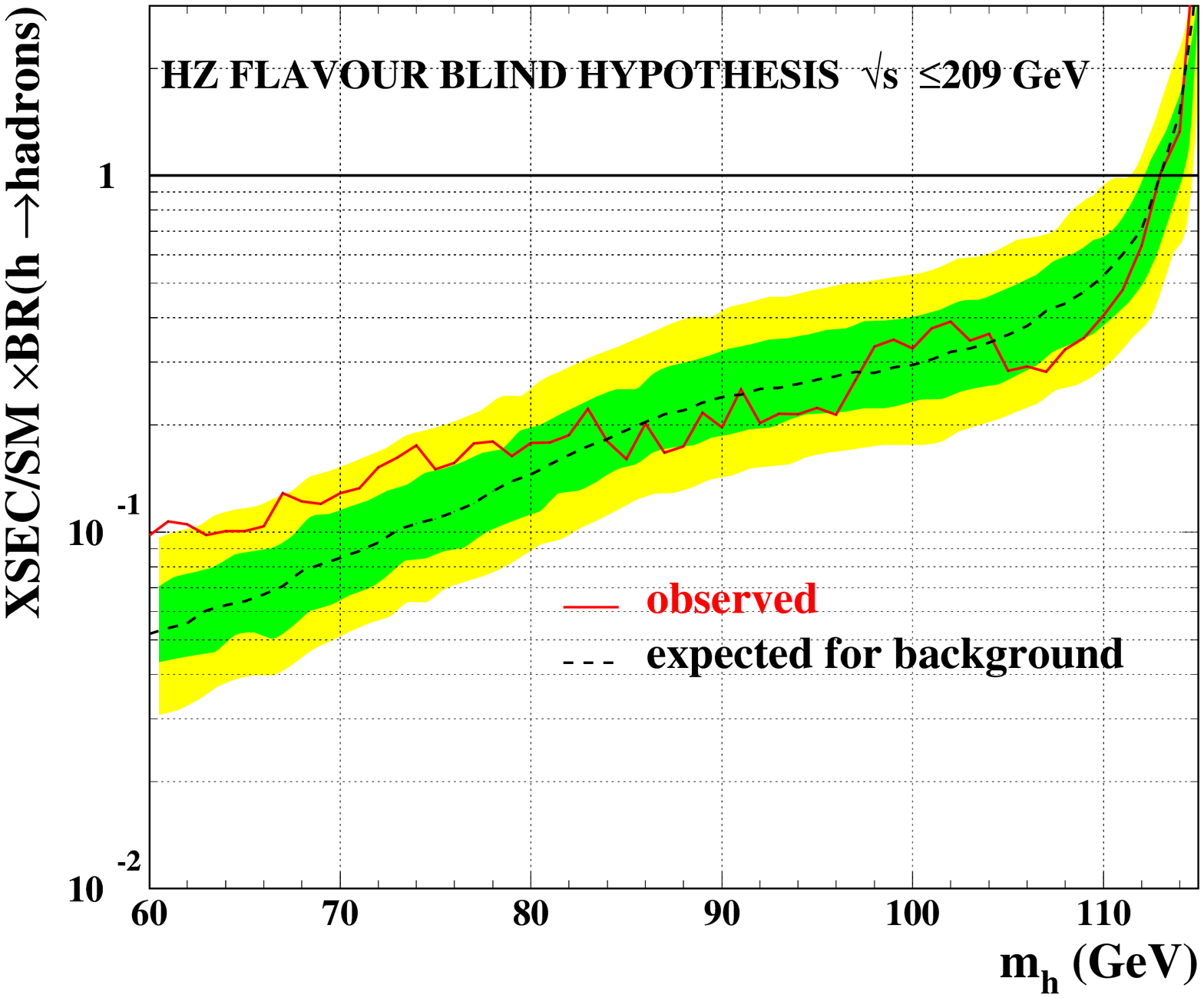,width=0.48\textwidth}
\vspace*{-0.3cm}
\caption[]{
Left:
The $CL_{\rm s}$ distribution for all hadronic and charged leptonic channels 
for flavor-independent hadronic Higgs boson decays, combining the data
collected by the four LEP experiments at energies up to 209~GeV.
The production cross section of the SM and Br$(\rm h\ra q\bar{q})=1$
is assumed.
The observed limit is given by the solid line and the expected limits 
are indicated by the dashed line and the shaded bands
($1\sigma$ and $2\sigma$).
The horizontal line at $CL_{\rm s}=0.05$ gives the limit at 95\% CL.
Right:
The 95\% CL bounds on the production cross section normalized 
to the SM cross section.
The observed limit is given by the solid line and the expected limits 
are indicated by the dashed line and the shaded bands 
($1\sigma$ and $2\sigma$).
\label{flavor-limit}}
\end{center}
\vspace*{-0.8cm}
\end{figure}

While in the SM the Higgs boson decays predominantly into b-quarks,
in many extensions such as the general two Higgs doublet model, the
b-quark~coupling~could~be~suppressed. 
Therefore, the search for
Higgs boson bremsstrahlung in\,the\,four-jet,\,two-jet\,and\,two-lepton, and
two-jet and missing-energy channels is generalized for flavor-independent
hadronic Higgs boson decays~\cite{flavor}.
Stringent limits are given in Fig.~\ref{flavor-limit}.
The expected limit is only 2.6 GeV below the SM limit, which is 
remarkable since b-tagging is\,a\,very\,important\,search\,tool.

\clearpage
\section{Photonic Higgs Boson Decays}
\begin{figure}[htb]
\vspace*{-0.9cm}
\begin{center}
\epsfig{figure=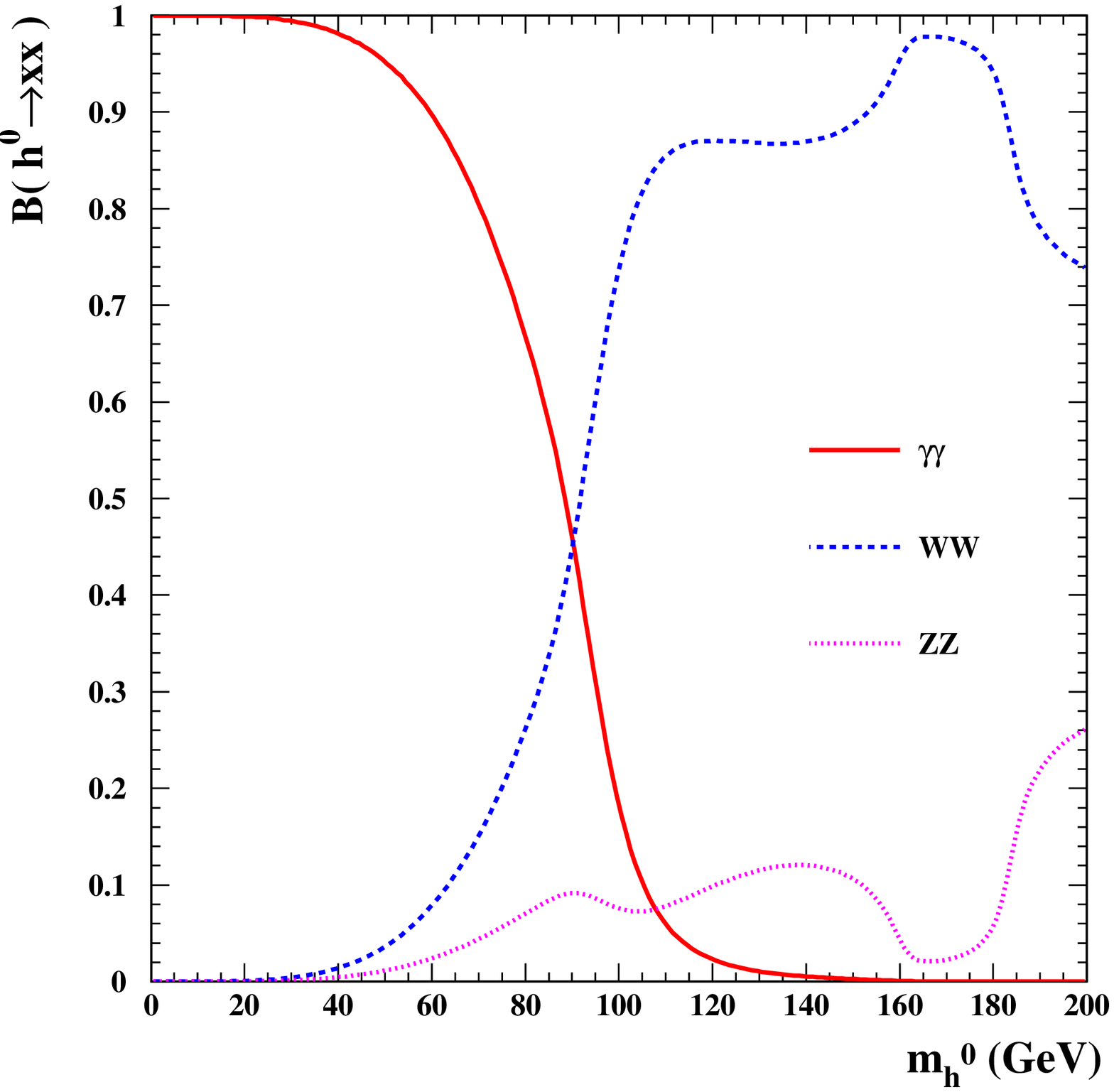,width=0.48\textwidth}
\epsfig{figure=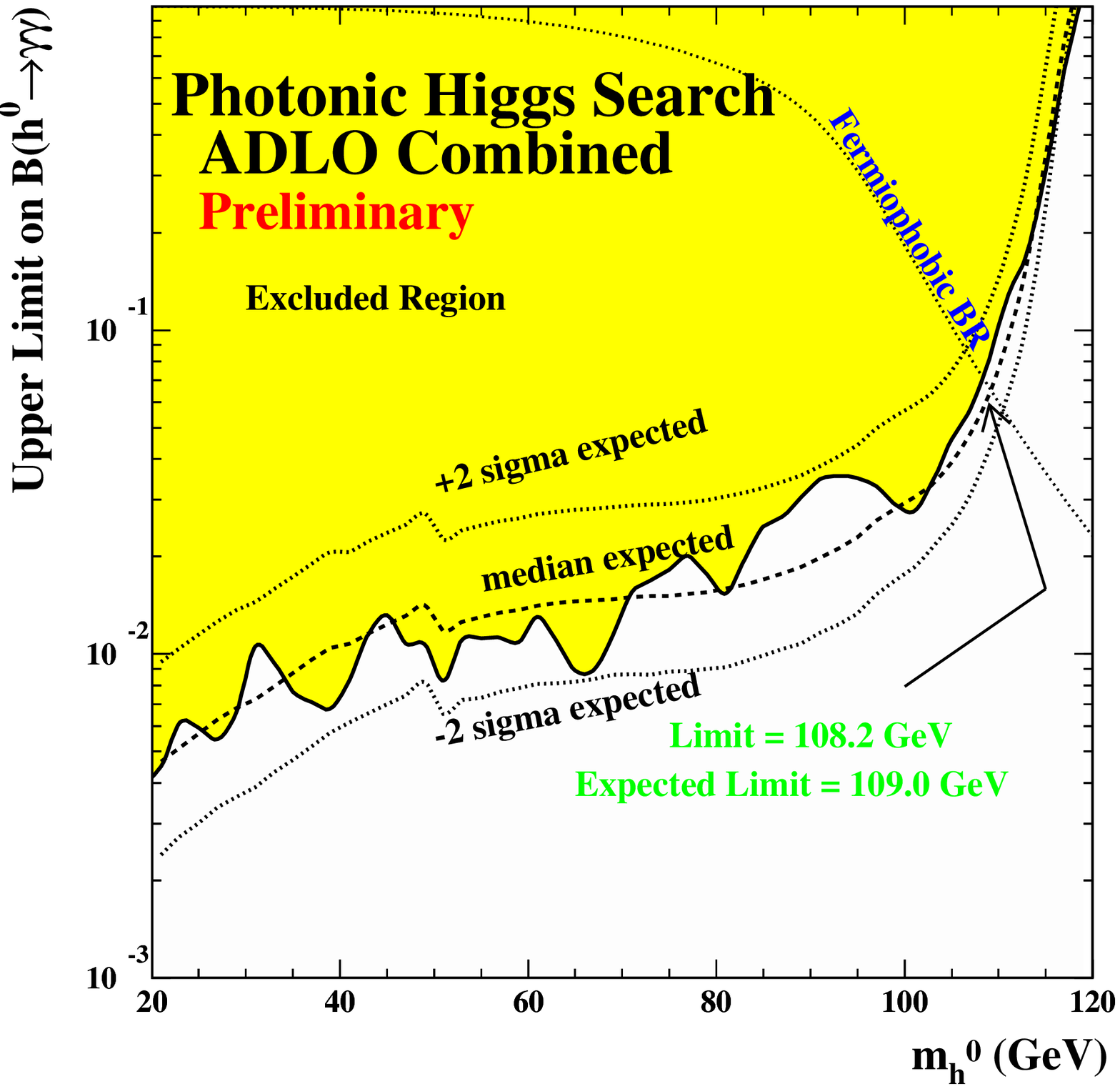,width=0.48\textwidth}
\vspace*{-0.6cm}
\caption[]{
Left:
Branching fraction of the fermiophobic benchmark model where only
the bosonic Higgs decay modes are allowed.
Right:
Limits at 95\% CL on the branching fraction $\rm h\ra\gamma\gamma$ assuming SM
production cross sections.
The observed limit is given by the solid line and the expected limits 
are indicated by the dashed line and the $2\sigma$ bands.
The dotted line gives the expected branching fraction in the
fermiophobic benchmark model.
\label{fermio-limit}}
\end{center}
\vspace*{-0.8cm}
\end{figure}

It is possible that the Higgs boson does not decay into
fermions. In this case the decays into $\gamma\gamma$, WW and ZZ 
are dominant. As a fermiophobic benchmark model, the production and
decays of the SM Higgs boson are assumed and all couplings to fermions 
are set to zero~\cite{fermio}.
Figure~\ref{fermio-limit} shows the expected Higgs boson branching
fractions and the mass limits from the search by all LEP experiments 
in the $\rm h\ra\gamma\gamma$ channel.

\section{Conclusions}
The combination of the complete data from the four LEP experiments
resulted in a large increase for the sensitivity of Higgs bosons.
The data shows a preference for a SM Higgs boson of 115.6~GeV.
Further small data excesses for Higgs boson pair-production and 
bremsstrahlung between 90 and 100~GeV allow the hypothesis that
h, A and H of the MSSM all have masses between 90 and 116~GeV.
Previously reported MSSM parameter combinations
from a general scan are supported by the complete data set.
The data is also consistent with the background-only hypothesis which
results in stringent mass limits for the SM Higgs boson, 
the neutral Higgs bosons of the MSSM,
charged Higgs bosons,
invisible Higgs boson decays,
flavour-independent hadronic Higgs boson decays,
and fermiophobic Higgs boson decays.
Table~\ref{tab:summary} summarizes these limits, and in addition
compares benchmark and scan limits in the MSSM.
\vspace*{-0.2cm}

\begin{table}[hp]
\begin{minipage}{0.43\textwidth}
\begin{tabular}{lrr}
Model             & Obs.        & Exp.\\\hline
SM                & 114.1       & 115.6 \\
MSSM ($m_{\rm h}$)&  91.0       &  94.6 \\
MSSM ($m_{\rm A}$)&  91.9       &  95.0 \\
$\rm H^+H^-$      &  78.6       &  78.8 \\
Invisible         & 114.3       & 113.6 \\
Flavor-independent& 112.9       & 113.0 \\
Fermiophobic      & 108.2       & 109.0 \\
                  &             & 
\end{tabular}
\end{minipage}
\begin{minipage}{0.52\textwidth}
\begin{tabular}{ll|rr|rr}
$\sqrt{s}$ (GeV)&Data& $m_{\rm h}^{\rm b}$ &$m_{\rm A}^{\rm b}$ &$m_{\rm h}^{\rm s}$ &$m_{\rm A}^{\rm s}$ \\\hline
91~\cite{91b,91s}   & L3      & 41.0 & none      & 25  & none \\
172~\cite{172b,172s}& DELPHI  & 59.5 & 51.0      & 30  & none \\
183~\cite{183b,183s}& DELPHI  & 74.4 & 75.2      & 67  & 75   \\
189~\cite{189b,189s}& DELPHI  & 82.6 & 84.1      & 75  & 78   \\
189~\cite{189bsopal}& OPAL&74.8& 76.5            & 72  & 76  \\
202~\cite{202bs}& DELPHI  & 85.9 & 86.5      & 85  & 86   \\
202~\cite{202ball,202sall}& LEP&88.3& 88.4      & 86  & 87   \\
209~\cite{209b,209s}& DELPHI  & 89.6 & 90.7      & 89  & 89   
\end{tabular}
\end{minipage}
\vspace*{-0.3cm}
\caption{\label{tab:summary}
Left: Observed and expected Higgs boson mass limits from complete LEP data in 
various models. Right: Benchmark (b) and scan (s) mass limits in the MSSM.
All limits are in GeV at 95\% CL.}
\vspace*{-0.4cm}
\end{table}

\clearpage
\section*{Acknowledgments}
\vspace*{-0.1cm}
I would like to thank the organizers of the conference
for their kind hospitality, and 
Pierre Lutz, Bill Murray and Alex Read for comments on the 
manuscript.

\end{document}